\documentclass[preprints,article,accept,moreauthors,pdftex,10pt,a4paper]{mdpi}

\graphicspath{{./Figures_July_2018/}}

\firstpage{1} 
\pubvolume{xx}
\issuenum{1}
\articlenumber{1}
\pubyear{2018}
\copyrightyear{2018}
\externaleditor{Academic Editor: name}
\history{Received: date; Accepted: date; Published: date}

\pdfoutput=1

\Title{Vortex dynamics in trabeculated embryonic ventricles}

\Author{Nicholas A. Battista $^{1}$*\orcidA{0000-0003-2437-0383}, Dylan R. Douglas $^{2,3}$, Andrea N. Lane $^{4}$, Leigh Ann Samsa $^{5}$, Jiandong Liu $^{6,7}$ and Laura A. Miller $^{2,3}$}

\AuthorNames{Nicholas A. Battista, Dylan R. Douglas, Andrea N. Lane, Leigh Ann Samsa, Jiandong Liu, and  and Laura A. Miller}

\address{%
$^{1}$ \quad Dept. of Mathematics and Statistics, 2000 Pennington Road, The College of New Jersey, Ewing Township, NJ 08628\\
$^{2}$ \quad Dept. of Biology, 3280, University of North Carolina, Chapel Hill, NC, 27599\\
$^{4}$ \quad Dept. of Biostatistics and Bioinformatics, Emory University, Atlanta, GA 30307\\
$^{3}$ \quad Dept. of Mathematics, CB 3250, University of North Carolina, Chapel Hill, NC, 27599\\
$^{5}$ \quad Dept. of Cell Biology and Physiology, University of North Carolina, Chapel Hill, NC 27599\\
$^{6}$ \quad McAllister Heart Institute, UNC School of Medicine, University of North Carolina, Chapel Hill, NC 27599\\
$^{7}$ \quad Dept. of Pathology and Laboratory Medicine, University of North Carolina, Chapel Hill, NC 27599}

\corres{battistn@tcnj.edu; Tel.: +1-716-517-6561}

\abstract{Proper heart morphogenesis requires a delicate balance between hemodynamic forces, myocardial activity, morphogen gradients, and epigenetic signaling, all of which are coupled with genetic regulatory networks. Recently both \textit{in vivo} and \textit{in silico} studies have tried to better understand hemodynamics at varying stages of veretebrate cardiogenesis. In particular, the intracardial hemodynamics during the onset of trabeculation is notably complex - the inertial and viscous fluid forces are approximately equal at this stage and small perturbations in morphology, scale, and steadiness of the flow can lead to significant changes in bulk flow structures, shear stress distributions, and chemical morphogen gradients. The immersed boundary method was used to solve the computational fluid dynamics problem involving fluid flow moving through the trabeculated ventricles of $72$, $80$, and $120$ hours post fertilization \textit{wild type} zebrafish embryos and \textit{ErbB2}-inhibited embryos at $7$ days post fertilization. An idealized trabeculated ventricular model was explored to map the bifurcations in flow structure that occur as a result of the unsteadiness of flow, trabeculae height, and fluid scale ($Re$). Vortex formation occurred in intertrabecular regions for biologically relevant parameter spaces, wherein flow velocities increased. This indicates that  trabecular morphology may alter intracardial flow patterns and hence ventricular shear stresses and morphogen gradients. A potential implication of this work is that the onset of vortical (disturbed) flows can upregulate Notch1 expression in endothelial cells \textit{in vitro} and hence impacts chamber morphogensis, valvulogenesis, and the formation of the trabeculae themselves.}


\begin{document}


%
%

%
%

\section{Introduction}
\label{Introduction}

Since the developing heart and embryo continue to function for some time in the absence of erythrocytes, it appears that the function of the early embryonic heart is not for the purpose of nutrient transport\cite{Burggren:2004,Patterson:2005}. Rather, recent work suggests that the heart's function is to aid in its own growth \cite{Hove:2003,Burggren:2004,Lindsey:2014,Santhanakrishnan:2011}. Two important roles for intracardial fluid dynamics in terms of proper cardiogenesis are to exert hemodynamic forces onto the ventricluar lining and to advect morphogens \cite{Icardo:1996,Patterson:2005}. These two fluid effects help regulate and drive organogenesis in developing embryos. Both shear stress and pressure may be key components to activating developmental regulatory networks by acting on cardiac cells \cite{Tarbell:2005} through a process called mechanotransduction. In this case, mechanical stimuli are transmitted by intracellular signalling pathways to the interior of the cell. Moreover increased receptor-ligand bond formation may appear near the endothelial lining in regions of higher vorticity \cite{Taylor:1996}, which gives rise to greater mixing of chemical morphogens. These chemicals may act as  epigenetic signals, which are then advected throughout the chamber \cite{Cartwright:09,Freund:12}. It is clear that irregular hemodynamics leads to cardiomyopathies or embryonic death \cite{Gruber:2004,Hove:2003,Stekelenburg:2008,Kowalski:2014,Midgett:2014}.




Prior to trabeculation, the endocardial ventricular cells are smooth and polygonal in shape. During the onset of trabeculation, several endocardial cells become elongated and some extend cellular projections. Moreover these cells appear slightly more depressed than the surrounding endocardial cells. The depressions progressively become deeper and wider such that the endocardial cells invaginate the cardiac jelly and extend toward the basal surface of the myocardium. Eventually myocardial cells separate due to the potent endocardial cell invasion, and definitive trabeculae are formed \cite{Icardo:1987}. Hence trabeculae are composed of both myocardial and endocardium components. 

Proper trabeculation requires well-coordinated cardiac contraction \cite{Samsa:2015} and is particularly sensitive to local changes in the fluid environment \cite{Granados:2012}. It is thought that the trabeculae may serve as mechanotranductive structures and alter intracardial flows in a way that regulates shear stress and mixing near the endocardium \cite{Malone:2007,Vedula:2017,Lee:2018}.  Even if subtle trabeculation irregularities were masked, cardiac defects would magnify over time because of their effect on morphogenetic processes. For example, zebrafish embryos designed to lack normal trabeculation (\textit{ErbB2}-inhibited) displayed severe cardiovascular defects including bradycardia (decreased heart rate), decreased fractional shortening, and impaired cardiac conduction \cite{Liu:2010}. Lack of trabeculae or irregularly formed trabeculae will cause irregular patterns of shear stresses. This in turn can cause dysfunctional myocardial activation patterns that are known to cause arrhythmias, abnormal fractional shortening, and even ventricullar fibrillation \cite{Reckova:2003}.



During the onset of trabeculation, the underlying fluid dynamics are particularly interesting due to the balance of inertial and viscous forces. The Reynolds number, $Re$, is a dimensionless number that describes the ratio of inertial to viscous forces in the fluid. It is given by $Re = (\rho U L)/\mu$, where $\mu$ and $\rho$ are the dynamic viscosity and density of embryonic blood, respectively, and $L$ and $U$ are characteristic length and velocity scales. The characteristic velocity is often chosen as the average or peak flow rate, while $L$ is often selected as the diameter of the chamber or vessel. When trabeculation begins as cardiac looping and ballooning progress, the $Re$ is approximately 1. At this fluid scale, a number of important fluid dynamic transitions can occur. One notable feature is the transition to vortical (disturbed) flow and hence changes in flow direction. This transition is sensitive to  the growing complex morphology, effective viscosity of the blood, and unsteadiness of the flow.

Disturbed blood flow patterns have been observed during heart development \cite{Liebling:2006}. Smoothly streaming (non-disturbed) blood flow induces higher wall shear stress (WSS), which inhibits endothelial cell activation. In contrast disturbed flow generally reduces WSS. Lower WSS helps stimulate adverse remodelling \cite{Chiu:2011,Morris18}. It has been shown that disturbed (vortical) flow patterns upregulate the expression of certain genes, such as Notch1, in endothelial cells  during development \cite{Jahnsen:2015}. As the heart undergoes dramatic morphological transformations, transitions to vortical and disturbed flow patterns may help guide morphogenesis through changing patterns in WSS or through other mechanotransductive mechanisms such as flow sensing through primary cilia\cite{HeidenKim:2006,Egorova:2011,Haack:2016}. Note that intracardial flows are both temporally and spatially varying such that the distribution of WSS is not uniform along the endothelium \cite{Vedula:2017,Lee:2018}. Hence mechanotransducers will exhibit different responses, leading to differentiated cellular behavior \cite{Koefoed:2014}. Furthermore, as the heart grows blood flows also increase \cite{Midgett:2015}. The formation of complex structures along the ventricle, like trabeculae, may provide regions where disturbed flow develops which could lead to higher kinetic energy dissipation. This energy dissipation may facilitate proper ventricle contractile function and trabecular organization \cite{Lee:2018}. 

Due to the complexity of the cardiogenesis and the challenges of measuring flow patterns precisely, computational fluid dynamics (CFD) has become a premier tool for resolving the flow in embryonic hearts \cite{DeGroff:2003,Taber:2007,Liu:2007,Santhanakrishnan:2009,Miller:2011,Buskohl:2012,Lee:2013,Battista:2016a,Battista:2017,Vedula:2017,Battista:2017b,Lee:2018}. For example, Liu \textit{et al.} \cite{Liu:2007} simulated flow through a three-dimensional model of a chick embryonic heart during stage HH21 (after about 3.5 days of incubation) at a maximum $Re$ of about 6.9. They found that vortices formed during the ejection phase near the inner curvature of the outflow tract. In 2013, Lee \textit{et al.} \cite{Lee:2013} performed 2D simulations of the developing zebrafish heart with moving cardiac walls. They found that unsteady vortices develop during atrial relaxation at 20-30 hpf and in both the atrium and ventricle at 110-120 hpf. 

More recently, Vedula \textit{et al.} \cite{Vedula:2017} and Lee \textit{et al.} \cite{Lee:2018} used light-sheet fluorescent microscopy and reconstructed a $4D$ moving ventricle to which they based their CFD model. They were able to quantify spatially- and temporally-varying WSS along trabceular ridges (trabeculae ``heads") and groves (``intertrabecular regions"). In particular Lee \textit{et al.} discovered that pulsatile shear-stresses developed along the ridges at 3 dpf in wildtype (WT) zebrafish embryos, while oscillatory shear-stresses (OSS) developed in the groves around 4 dpf \cite{Lee:2018}. Around 4 dpf vortical flow patterns may be present within the intertrabecular spaces. Moreover, OSS were found to be substantially less at the trabecular heads, suggesting that OSS may be a possible regulatory control during cardiogenesis \cite{Vedula:2017,Lee:2018}. They also investigated differences in WSS between wildtype and mutant zebrafish hearts. The mutants they considered were \textit{ErbB2}-inhibited zebrafish (suppresses trabeculation), \textit{gata1a} morpholinos (lowers blood viscosity), and \textit{wea} mutants (lower cardiac contractility). They found that total WSS was comparable in the WT and ErbB2-inhibited zebrafish; however, the \textit{gata1a} morpholinos and \textit{wea} mutants expressed significantly less total WSS. Another study, Battista \textit{et al.} \cite{Battista:2017}, found that trabeculae morphology has a significant effect on intertrabecular vortex formation, as does the presence of hematocrit and fluid scale. However, their study did not include an analysis of WSS, but instead referred to the tangential, normal, and total force magnitudes as potential proxies for WSS, although they observed similar trends to the spatially-averaged WSS over the course of a heart cycle.

The numerical work described above and \textit{in vivo} measurements of blood flow in embryonic hearts \cite{Hove:2003,Vennemann:2006} supports that vortex formation is sensitive to changes in $Re$, morphology, and unsteadiness of the flow. Santhanakrishnan \textit{et al.} \cite{Santhanakrishnan:2009} used a combination of CFD and flow visualization in dynamically scaled physical models to describe the fluid dynamic transitions that occur as the chambers balloon, the endocardial cushions grow, and the overall scale of the heart increases. They found that the formation of intracardial vortices depends upon the height of the endocardial cushions, the depth of the chambers, and the $Re$. Their study only considered steady flows in an idealized two-dimensional chamber geometry with smooth, stationary walls.

In this paper, we present complementary studies to both Santhanakrishnan \textit{et al.} \cite{Santhanakrishnan:2009} and Lee \textit{et al.} \cite{Lee:2018} with the goal of revealing the bifurcations in flow structures that occur as a result of the unsteadiness of the flow, trabeculae height, and $Re$. First we investigate the differences in the cardiac fluid dynamics between WT and \textit{ErbB2}-inhibited (namely \textit{ErbB2$^{st61}$} and \textit{ErbB2$^{st50}$}) mutants, to explore how vortex formation in the intertrabecular regions is sensitive to differences in morphology. We quantify the intertrabecular flow patterns mentioned (but not shown) in \textit{Lee et al.} \cite{Lee:2018}. Next, we use an idealized geometry, based upon that of Santhanakrishnan \textit{et al.} \cite{Santhanakrishnan:2009}, to systematically sweep a parameter space consisting of trabeculae size, fluid scale, and unsteady flow effects to quantify fluid dynamics transitions.




%
%

%
%

\section{Methods}
\label{numerical_section}

Two-dimensional computational fluid dynamics (CFD) simulations were used to quantify the flow fields within a biologically-realistic and an idealized model of a trabeculated ventricle of the zebrafish embryonic heart. We will discuss the model geometry construction below. For a detailed discussion on the numerical method used to solve the fluid-structure problem, see Appendix \ref{app:IB_Method}.


%
%

\subsection{Embryonic Zebrafish Model Geometry}
\label{sec:bio_geometry}

We estimated the structure of trabeculae from \textit{in-vivo} image data taken from embryonic zebrafish for the purpose of simulating flow through realistic ventricle geometries. Five images of stained cross sections of both wildtype and mutant embryonic zebrafish were taken from Liu \textit{et al.} \cite{Liu:2010}. The mutant and transgenic lines used were \textit{ErbB2$^{st61}$} and \textit{ErbB2$^{st50}$}. Figures from the original paper were cropped out and manually traced by recording pixel locations along the boundary with the open-source Python package Argus \cite{Jackson:2016}. Smooth-spline approximations of the trabeculated ventricle walls were generated and sampled to generate a finite-difference mesh.  

\begin{figure}
\centering
\includegraphics[width=0.975\textwidth]{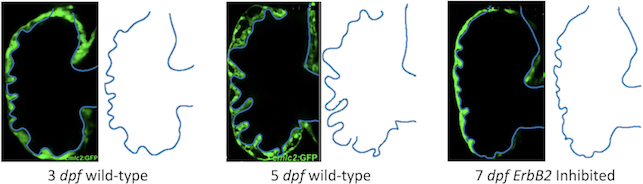}
\caption{The extracted geometries from trabeculated ventricles at different stages of development from Liu\textit{et al.} \cite{Liu:2010}. There are two cases shown for WT zebrafish at 3 and 5 \textit{dpf} and the geometry taken from an \textit{ErbB2}-inhibited zebrafish at 7 \textit{dpf}.}
\label{Bio_Geometry}
\end{figure}

Figure \ref{Bio_Geometry} shows the computational geometries extracted from microscopy images \cite{Liu:2010}. Note that the images were acquired using a Nikon Te-2000u microscope at a rate of 250 frames per second with a high-speed CMOS camera (MiCam Ultima, SciMedia). The geometries chosen for this study were of wild-type embryos at 3 and 5 \textit{dpf} and an \textit{ErbB2} inhibited embryo at 7 \textit{dpf}. Note that at 5 \textit{dpf} \textit{ErbB2}-inhibited embryos show little to no signs of trabeculation \cite{Liu:2010}; however at 7 \textit{dpf} there is the onset of trabeculation, as shown above. Furthermore, we considered other wild-type embryos at 80 \textit{hpf} and 5 \textit{dpf} to span biological variation, see Figure \ref{Bio_Geometry_App} in Appendix \ref{app:bio_geo} for their geometries.

%
%

\begin{figure}
\centering
\includegraphics[width=0.925\textwidth]{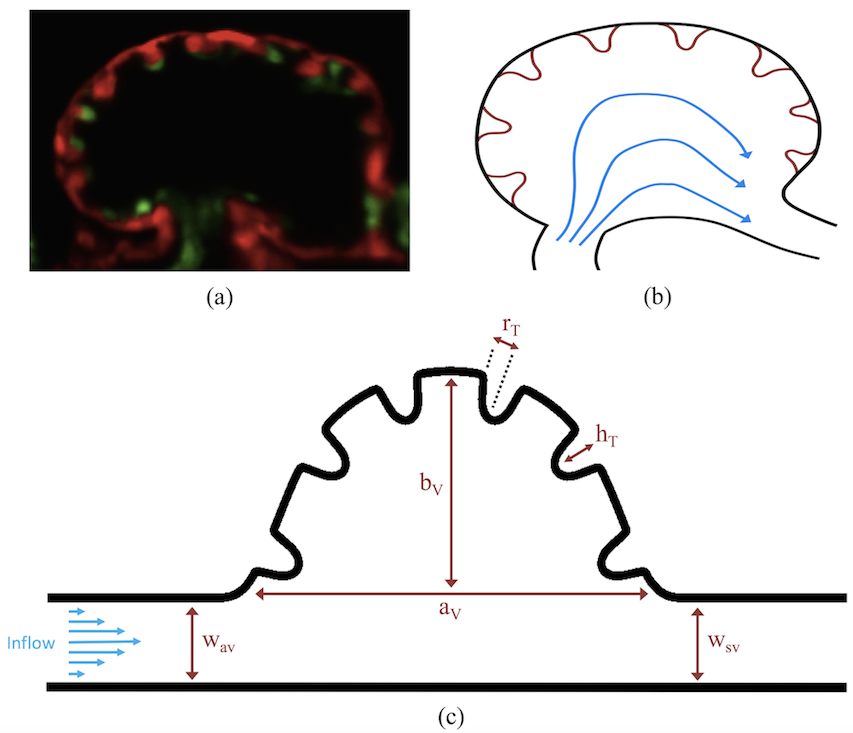}
\caption{(a) A microscopy image of an embryonic zebrafish's trabeculated ventricle at 4 \textit{dpf} from Liu \textit{et al.} \cite{Liu:2010}. This snapshot was taken immediately before the systolic phase. The protrusions into the ventricle are trabeculae. The image was taken from Tg(cmlc2:dsRed)s879; Tg(flk1:mcherry)s843 embryos expressing fluorescent proteins that label the myocardium and endocardium, respectively \cite{Liu:2010}. (b) Simplified diagram showing the basic idea behind our idealized geometry. Blood flows from the atrio-ventricular canal into the ventricle and then proceeds into the bulbus arteriosus. (c) Further idealization of the computational model which is now flattened. The geometric parameters are as follows: $a_V$ and $b_V$ are the semi-major and semi-minor axis of the elliptical chamber, $h_T$ and $r_T$ are the height and radii of the trabeculae, and $w_{AV}$ and $w_{SV}$ are the widths of the AV canal and sinus venosus respectively.}
\label{Model_Geometry}
\end{figure}

To ensure the simulations were conducted at the appropriate fluid scale, we computed the biologically relevant $Re$ to encapsulate the correct fluid dynamic regime. For the $Re$ within the ventricle of a $4\ dpf$ wild type zebrafish, the characteristic velocity, $V_{zf}$ was taken as the average of the minimum and maximum velocity measured \textit{in vivo}, and the characteristic length, $L_{zf}$, was taken as the width of the ventricle from Figure \ref{Model_Geometry}. The $Re$ was then calculated as 
\begin{equation}
Re = \frac{\rho_{zf} L_{zf} V_{zf}}{\mu_{zf}} = 1.07,
\end{equation}
\noindent where $V_{zf} = 0.75$ $cm/s$ \cite{Hove:2003}, $\rho_{zf} = 1025$ $kg/m^3$ \cite{Santhanakrishnan:2011}, $\mu_{zf} = 0.0015$ $kg / (m s)$ \cite{Mohammed:2011}, and $L_{zf} = 208$ $\mu m$. The characteristic frequency was chosen as $f = 3.95\ beats/s$ \cite{Malone:2007}. The dimensionless frequency may then be calculated as
\begin{equation}
\tilde{f}=\frac{L_{zf}}{V_{zf}} f_{zf} = 0.11.
\end{equation}

The simulations using the biologically realistic geometry were performed at $Re\sim1$. The length scale was taken directly from the images of the embryos and the velocity, frequency, and kinematic viscosity (the ratio of the fluid's dynamic viscosity to density) were taken from the literature as described above. 

\begin{figure}
\centering
\includegraphics[width=0.325\textwidth]{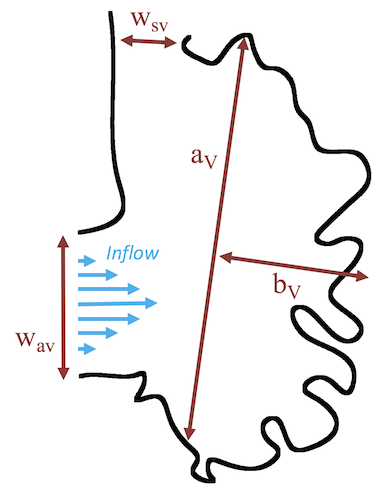}
\caption{Schematic diagram showing the parameters for the biologically realistic simulations. Note that $a_V$ is the characteristic length for selecting the correct $Re$.}
\label{Bio_Comp_Geometry}
\end{figure}

Figure \ref{Bio_Comp_Geometry} shows how the geometric parameters were measured from the reconstructed geometries. Note that although there is geometric variation between each case, the parameters are labeled consistently. We do not, however, vary the $Re$ for the simulations that use the realistic geometry. Note that we do vary $Re$ for the idealized trabeculated ventricle geometry described in Section \ref{Geometry}.

%
%

\subsection{Idealized Model Geometry}
\label{Geometry}

A simplified two dimensional geometry of a 4 \textit{dpf} zebrafish's trabeculated ventricle was constructed using Figure \ref{Model_Geometry}a, from Liu \textit{et al.} \cite{Liu:2010}. The ventricle was idealized as a half ellipse, with semi-major axis $a_V$ and semi-minor axis $b_V$. It is tangentially laid within a channel, which emulates a cavity-flow geometry. The channel models the atrioventricular canal (AV canal), with width $w_{AV}$ set equal to the width of the sinus venosus (SV), $w_{SV}$. Six equally-spaced trabeculae were aligned within the ventricle. The idealized model geometry is illustrated in Figure \ref{Model_Geometry}c. 

The idealized trabeculae geometry was modeled using the following perturbed Gaussian-like function,
\begin{equation}
\label{TrabGeometry} T(x) = h_T \left(1 - \left(\frac{x}{r_T}\right)^2 \right) e^{-\left(\frac{x}{0.7r_T} \right)^8 },
\end{equation}
where $r_T$ and $h_T$ are the radii and height of each trabecula, respectively. The full idealized geometry can be seen in Figure \ref{Model_Geometry}c.

The geometric model parameters are given in Table \ref{Geometric parameters of the computational model}. All parameters were scaled from measurements taken from Figure \ref{Model_Geometry}a. The parameters describing the ventricle were held constant and are given as the chamber height, $b_V$, chamber width, $a_V$,  and width of the AV canal and SV, $w_{AV}$ and $w_{SV}$ respectively. Note that the radii of the trabeculae, $r_T$, was held constant in all numerical simulations, while the height of the trabeculae, $h_T$, was varied.

\begin{table}[H]
\centering
\begin{tabular}{|c|c|}
\hline
Parameter & Value \\
\hline
$a_V$ & 1.0 \\
$b_V$ & 0.8 \\
$w_{AV}$ & 0.8 \\
$w_{SV}$ & 0.8 \\
$r_T$ & 0.10 \\
$\frac{h_T}{b_V}$ & \{0, 0.02, 0.04, \ldots, 0.16\}\\
\hline
\end{tabular}
\caption{Table of geometric parameters used in the idealized numerical model. The height of trabeculae, $h_T$, were varied for numerical experiments.}
\label{Geometric parameters of the computational model}
\end{table}

The temporal parameter values were chosen to keep the dimensionless frequency fixed at 0.10 for the pulsatile simulations, analogous to the model presented in Section \ref{sec:bio_geometry}. The $Re$ was varied by changing the kinematic viscosity, $\nu = \mu / \rho$. The computational parameters are found in Table \ref{BCs}.  For the simulations, the $Re_{sim}$ is calculated using a characteristic length of $w_{AV}$ and a characteristic velocity set to $V_{in}$ (steady inflow) or $\frac{1}{2} V_{in}$ (pulsatile inflow). The simulations were performed for $Re_{sim} =  {0.01, 0.05, 0.1, 0.5, 1, 5, 10, 20, 30, 40, 50, 100}$. The stiffness of the tether springs were chosen the minimize the deformations of the boundary to below $1\%$ of the chamber diameter. This approach is used in the immersed boundary method to describe a nearly rigid boundary.

Note that we consider $Re$ two orders of magnitude higher and lower than the relevant $Re\sim1$ for a $4\ dpf$ zebrafish heart. The reason for this is two fold: 1) to map out the parameter space outside of the biologically relevant range and 2) to provide insight into fluid flows in other types of trabeculated hearts, such as those of some  invertebrates \cite{McGaw:2002,Wilbur:2013}.

\section{Results}

Below we present the flow patterns and velocities for both biologically realistic and idealized $2D$ models of trabeculated ventricles. The cases with biologically realistic geometries were of WT zebrafish and an \textit{ErbB2}-inhibited mutant to contrast the intracardial and intertrabecular fluid dynamics of an embryonic zebrafish heart during development. The idealized geometry was used to systematically sweep over a parameter space to describe transitions in flow patterns for pulsatile flows, changes in trabecular height, and Reynolds Number $Re$. In the idealized geometry case, the $Re$ was varied from 0.01 to 100, and the trabecular heights were varied from zero to twice the biologically relevant height. We also quantified flow for both steady and pulsatile cases.

Streamlines were used to show the path that a passive particle would take in the flow. The streamline graphs were generated using the VisIt visualization software \cite{HPV:VisIt}. The streamlines are drawn by making a contour map of the stream function, since the stream function is constant along the streamline. The stream function, $\psi({\bf{x}}, t)$, in 2-D is defined by the following equations:
\begin{eqnarray}
u({\bf{x}},t)=\frac{\partial \psi({\bf{x}},t)}{\partial y}\\
v({\bf{x}},t)=-\frac{\partial \psi({\bf{x}},t)}{\partial x}
\end{eqnarray}

\noindent The streamline colors correspond to smooth, streaming flow (blue) and vortical flow (orange).

%
%

\subsection{Steady flow through an embryonic zebrafish heart}
\label{results:steady_bio}

\begin{figure}
\centering
\includegraphics[width=0.975\textwidth]{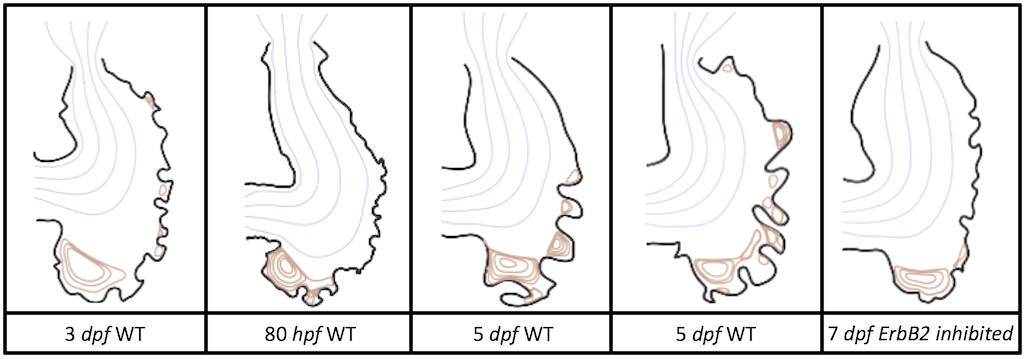}
\caption{Streamlines showing the direction of steady flow through realistic trabeculated ventricular geometries.}
\label{fig:real_Steady}
\end{figure}

Figure \ref{fig:real_Steady} gives the flow field streamlines for the steady inflow cases with biologically realistic geometries at $Re\sim1$. Parabolic inflow wwas used that accelerates from rest to a constant velocity, as detailed in Appendix \ref{app:IB_Method}. The simulation data is given once the flow has reached its steady state. Results are shown for a 3 \textit{dpf}, 80 \textit{hpf}, and two 5 \textit{dpf} wild-type embryos as well as an \textit{ErbB2}-inhibited embryo at 7 \textit{dpf}. In all cases, vortex formation occurred in the intertrabecular regions along the side opposite that to the sinus venosus (SV) and was within the well pronounced intertrabecular grooves. No intracardial vortices were observed and the flow smoothly steams from the atrioventricular canal to the sinus venosus. 

\begin{figure}
\centering
\includegraphics[width=0.975\textwidth]{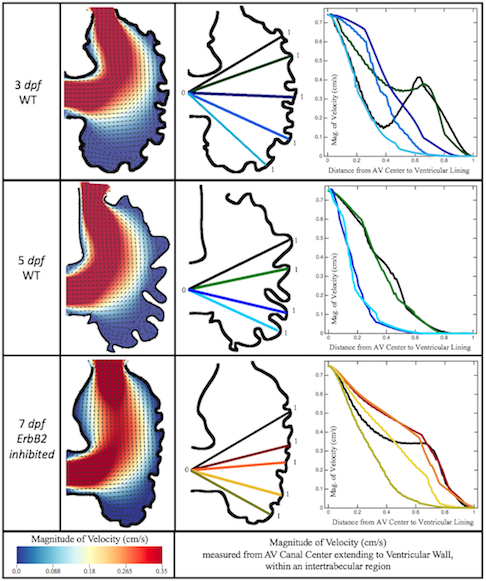}
\caption{Simulation results for a 3 and 5 $dpf$ WT and 7 \textit{dpf ErbB2} inhibited zebrafish. The magnitude of velocity (cm/s) is given by the colormap, and the arrows show the direction of flow once steady state is reached (left). The panels on the right show the magnitude of velocity along lines drawn from the center of the AV canal to the ventricular lining between the trabeculae. Note that these lines are shown in the middle panels. }
\label{fig:real_Steady_uMag_uVec_combined}
\end{figure}

Figure \ref{fig:real_Steady_uMag_uVec_combined} illustrates that in the regions of significant vortex formation, e.g., in the intertrabecular regions, velocities are much lower than in the intracardial region of the ventricle. In all cases the velocities on the side opposite the SV experience much faster velocity decay towards the ventricle lining, whereas regions opposite the AV canal experience slower decay. Note that in the 3 \textit{dpf} and 7 \textit{dpf ErbB2} inhibited cases, the flow velocity along the line drawn from the AV canal to the intertrabecular region that is closest to the SV decreases in magnitude before it increases again and finally decays to zero near the cardiac wall. This is due to the fact that the velocity is measured close to the ventricular wall that extends from the AV canal. Flow velocities measured between the trabeculae are significantly lower than those measured in the within the middle of the chamber.

%
%

\subsection{Pulsatile flow through an embryonic zebrafish heart}

In this subsection we show the results of the pulsatile inflow simulations where  biologically realistic geometries are used at $Re\sim1$ and the frequencies are varied. Temporal snapshots of the streamlines are presented in Figure \ref{fig:real_Pulsatile}. The pulsatile inflow condition is detailed in Appendix \ref{app:IB_Method}. All data shown is taken from the last pulse cycle such that periodic steady state has been reached. The geometries simulated include a 3 \textit{dpf}, a 80 \textit{hpf}, and two 5 \textit{dpf} wild-type embryos as well as an \textit{ErbB2} inhibited embryo at 7 \textit{dpf}. 

\begin{figure}[H]
\centering
\includegraphics[width=0.99\textwidth]{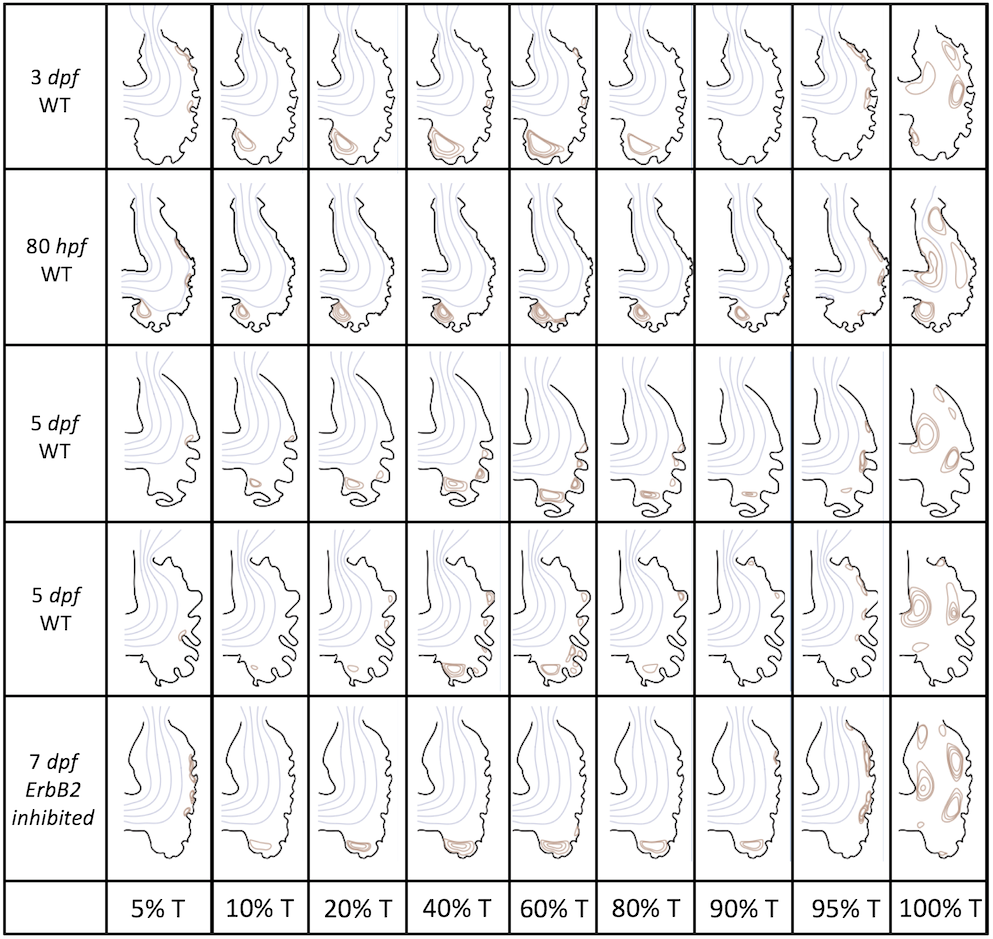}
\caption{Temporal snapshots of the streamlines showing pulsatile flow within the realistic trabeculated ventricles of embryonic zebrafish during different stages of development.}
\label{fig:real_Pulsatile}
\end{figure}

Vortices formed in all cases within the intertrabecular regions along the side opposite that to the sinus venosus, similar to the steady cases illustrated in Figure \ref{fig:real_Steady}. The vortices are, however, dynamic; they change shape and size during a single pulsation cycle. Some additional vortices appear in the intertrabecular regions that were not present in the steady case. For example, consider the 80 \textit{hpf} and \textit{ErbB2}-inhibited cases where vortices appear on the side opposite the AV canal. Furthermore, large intracardial vortices appear between pulsation cycles in all cases when there is minimal inflow. 

\begin{figure}
\centering
\includegraphics[width=0.975\textwidth]{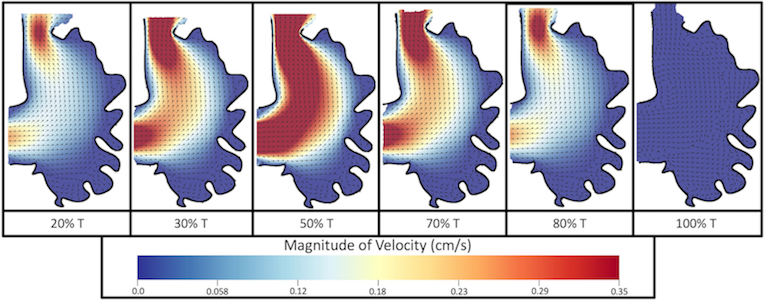}
\caption{Temporal snapshots taken during on pulse showing the flow simulated within a 5 \textit{dpf} WT ventricle. The magnitude of velocity (cm/s) is given by the colormap and the direction of flow is given by the arrows.}
\label{fig:real_Pulsatile_5dpf}
\end{figure}

\begin{figure}
\centering
\includegraphics[width=0.975\textwidth]{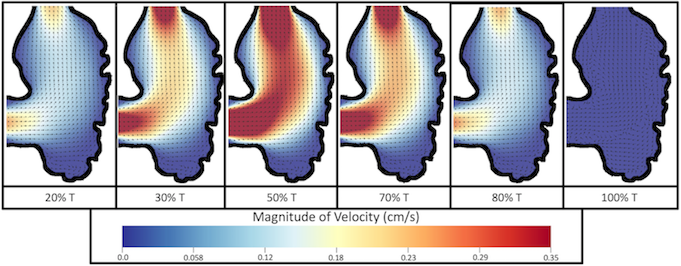}
\caption{Temporal snapshots of the flow during one pulse cycle taken from an \textit{ErbB2}-inhibited based geometry at 7 \textit{dpf}. The magnitude of velocity (cm/s) is shown by the colormap along with the velocity vectors.}
\label{fig:real_Pulsatile_ErbB2_7dpf}
\end{figure}

Figures \ref{fig:real_Pulsatile_5dpf} and \ref{fig:real_Pulsatile_ErbB2_7dpf} give temporal snapshots for a 5 \textit{dpf} WT and \textit{ErbB2}-inhibited cases, respectively. The colormap illustrates the magnitude of velocity (cm/s) with accompanying velocity vector fields. The velocity field is consistent with Figure \ref{fig:real_Pulsatile} and shows vortex formation. These results also illustrate the spatial gradient in velocity within the intracardial to intertrabecular regions. It is clear that although the fastest flow moves towards the middle of the chamber, the velocity significantly tapers off by the time it reaches the ventricular lining. 

\begin{figure}
\centering
\includegraphics[width=0.975\textwidth]{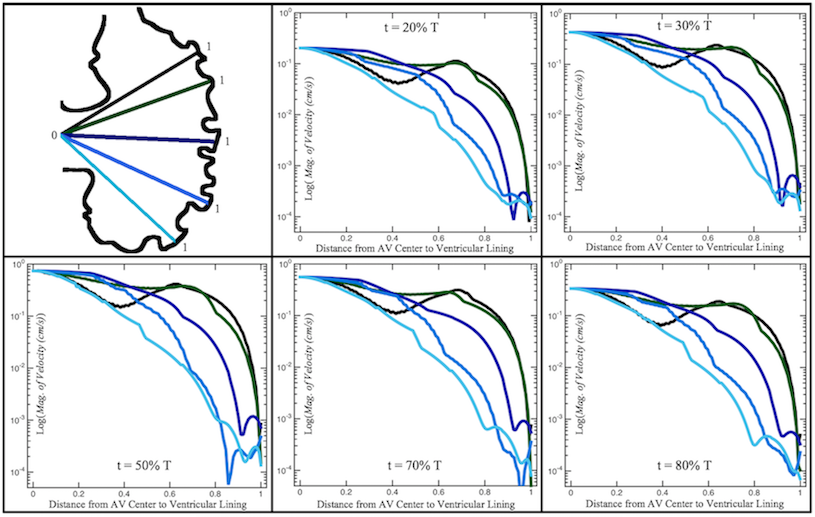}
\caption{Plots showing how the log magnitude of velocity decays from the center of the AV canal to the ventricle wall for five different lines across the chamber using a 3 \textit{dpf} WT embryo's geometry.}
\label{fig:real_Pulsatile_3dpf_LoguMag_Data}
\end{figure}

\begin{figure}
\centering
\includegraphics[width=0.975\textwidth]{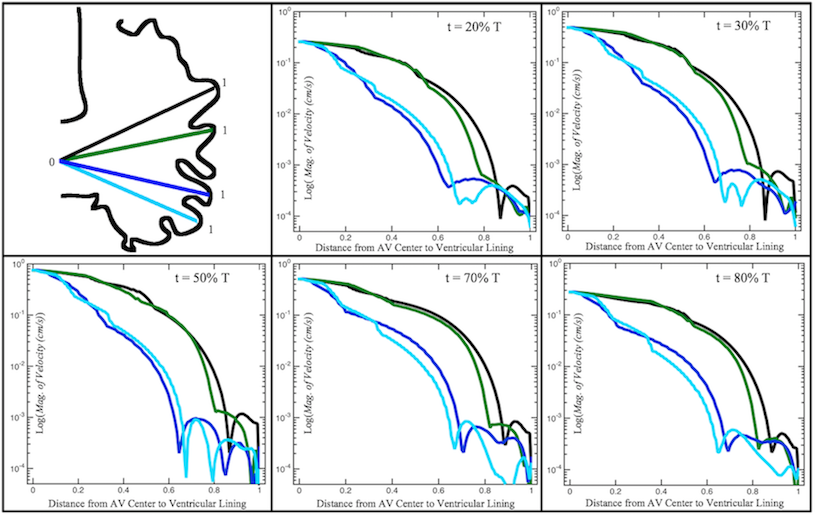}
\caption{Plots showing how the log magnitude of velocity decays from the center of the AV canal to the ventricle wall for four different lines across the chamber using a 5 \textit{dpf} WT embryo's geometry.}
\label{fig:real_Pulsatile_5dpf_LoguMag_Data}
\end{figure}

\begin{figure}
\centering
\includegraphics[width=0.975\textwidth]{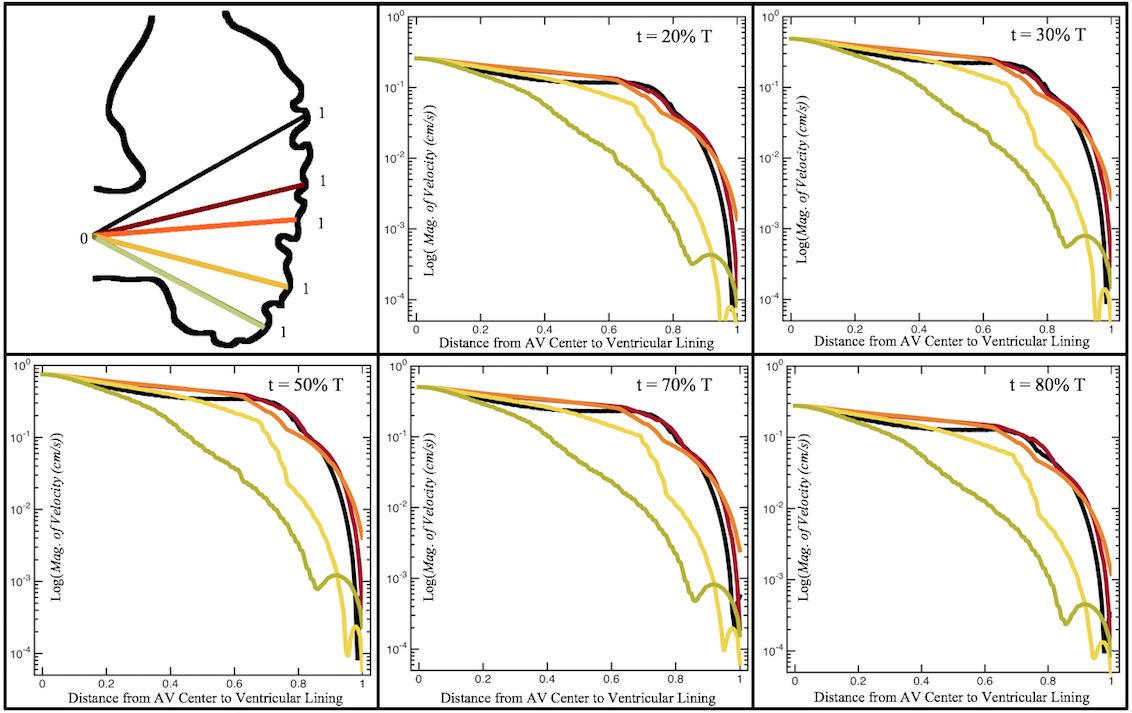}
\caption{Plots showing how the log magnitude of velocity decays from the center of the AV canal to the ventricle wall for five different lines across the chamber using a 7 \textit{dpf ErbB2}-inhibited embryo's geometry.}
\label{fig:real_Pulsatile_7dpf_ErbB2_LoguMag_Data}
\end{figure}

Figures \ref{fig:real_Pulsatile_3dpf_LoguMag_Data}, \ref{fig:real_Pulsatile_5dpf_LoguMag_Data}, and \ref{fig:real_Pulsatile_7dpf_ErbB2_LoguMag_Data} illustrate the magnitude of the velocity along lines drawn from the center of the AV canal to the intratrabecular regions for cases of a 3 and 5 \textit{dpf} WT and 7 \textit{dpf ErbB2}-inhibited zebrafish geometries.  Appendix \ref{app:bio_results} presents this data in a non-logarithmic scale, to complement the data given in Figure \ref{fig:real_Steady_uMag_uVec_combined} in Section \ref{results:steady_bio}. Note that the 3 \textit{dpf} WT and 7 \textit{dpf ErbB2}-inhibited zebrafish exhibit smaller trabeculae than the 5 \textit{dpf} WT case. 

As expected in all cases, peak flows occur within the chamber and decay as one moves toward the cardiac wall. Even in the steady inflow cases, the velocity decay is similar. The velocity decreases close to geometrically towards the intertrabecular region on the opposite side to the SV for the smaller trabeculae (3 \textit{dpf} WT and 7 \textit{dpf ErbB2}-inhibited). For the 3 \textit{dpf} WT embryos in regions with pronounced trabeculation, the velocity decay does not monotonically decrease; it increases slightly within the trabeculae before decreasing to zero as one moves towards the endocardium. Similar non-monotonic behavior is seen for some intratrabecular regions farther from the SV in the 3 \textit{dpf} WT and 7 \textit{dpf ErbB2}-inhibited embryos. We also note that the magnitude of flow decreases much more rapidly for the 5 \textit{dpf} WT with pronounced trabeculation, indicating lower WSS within the intratrabecular regions.


%
%
%
%
%
%

\subsection{Steady Flow through Idealized Trabeculated Chambers}
\label{sec:ideal_steady}


\begin{figure}
\centering
\includegraphics[width=0.925\textwidth]{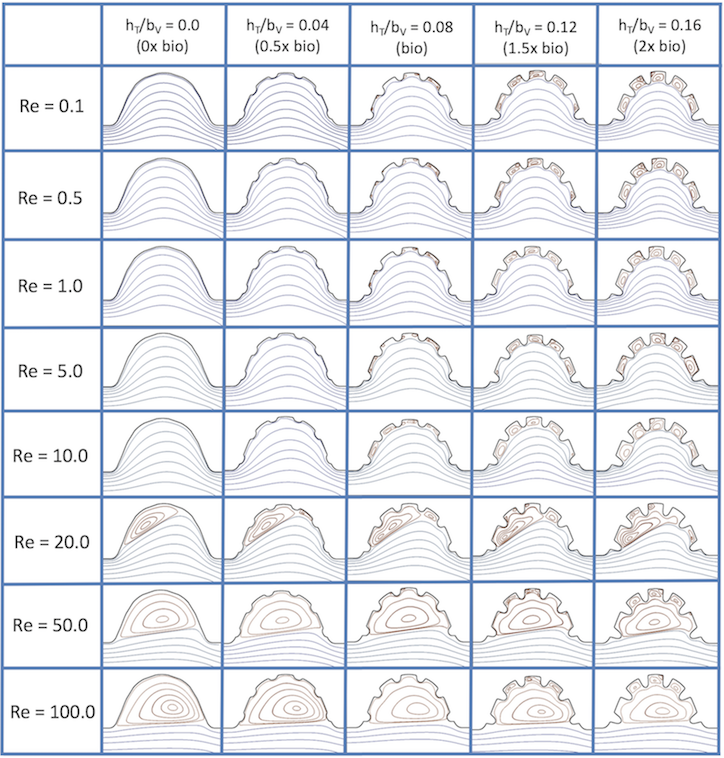}
\caption{Streamline analysis performed for the case of steady flow into the trabeculated ventricle of a zebrafish at 4 \textit{dpf} for varying $Re$ and trabeculae heights.}
\label{SteadyFlow_Results}
\end{figure}

Figure \ref{SteadyFlow_Results} shows the flow field streamlines for the case of steady flow through an idealized trabeculated embryonic ventricle. The inflow condition is detailed in Appendix \ref{app:IB_Method}. The numerical simulations span five orders of magnitude of $Re$, varying from 0.01 to 100, while trabeculae heights were set to $0 \leq \frac{h_T}{b_V} \leq 0.16$. Note that the biologically relevant case is  $\frac{h_T}{b_V} = 0.08$

In the case of no trabeculae (left column), we find vortex formation only occurs for $Re \geq 15$, in agreement with the findings of \cite{Santhanakrishnan:2009}.  Moreover this result appears consistent with fluid dynamics literature on transitions to vortical flow via a channel with an expanded region \cite{Shen:1985,Mizushima:1996}. Shen et al. 1985 and Mizushima et al. 1996 both use rectangular cavities; however, similar transitions to vortical flow occur at consistent $Re$. For $Re\leq 10$, the flow bends around the cavity and no flow separation occurs. As $Re$ is increased to $20$, flow reversal occurs and a closed vortex is present along the left side of the cavity. The stagnation point is located between the orange and blue streamlines. To the left of this stagnation point, the flow moves along the endocardium from the right to left. To the right of the stagnation point, the flow moves right to left.  As $Re$ is further increased, the stagnation point moves to the right, and the intracardial vortex becomes larger until it becomes as large as the cavity itself for $Re\sim100$.

When half-size biologically relevant trabeculae are introduced into the model ($\frac{h_T}{b_V}=0.04$), similar flow fields emerge for the case of $Re\leq10$. Although geometric perturbations now exist along the cavity lining, no flow separation occurs, whether intracardially or intertrabecularly. For $Re=20$ we see a similar intracardial vortex to the case without trabeculae; however, it is also seen to weave along regions with trabeculae. Furthermore there is an emergence of an independent closed vortex along the right side between two trabeculae. For $Re\geq50$, we find the presence of one large intracardial vortex wrapping around each trabeculae.  

For biologically relevant trabeculae heights, there are closed intertrabecular vortices for $Re$ as low as 0.1, while no intracardial vortices are present at these lower $Re$. Shen et al. 1985 \cite{Shen:1985} saw a similar phenemenon with the formation of two vortices in the corners of their rectangular cavity. This is consistent with the formation of vortices near the only bottom of the trabeculae. On the other hand, interestingly, not all intertrabecular regions have closed vortices. As $Re$ is further increased from $Re=5$ to $Re=10$, the intertrabecular vortices grow in size. As in previous cases, a larger intracardial vortex forms at $Re=20$. On the left hand side of the cavity, there is smooth flow from left to right around the trabeculae. On the right side of the cavity, independent closed vortices form between the trabeculae, and the flow is from right to left. For $Re\geq 50$, a large intracardial vortex forms and no intertrabecular vortices persist.

For trabeculae heights higher than the biologically relevant range, there exist intertrabecular vortices for $Re$ as low as $Re=0.1$; however, compared to the previous biologically relevant case, there are vortices between every adjacent pair of trabeculae. Moreover, because the trabeculae extend further into the ventricular cavity, these vortices are larger than in previous cases. Intracardial vortices do not develop until $Re\geq 20$, where there is the presence of one large intracardial vortex on the left side of the cavity. When $Re=20$ and $\frac{h_T}{b_V}=0.12$, the intracardial vortex only wraps itself around the first four trabeculae with flow moving from left to right. A single intertrabecular vortex forms in the fourth trabecular valley. When $Re=20$ and $\frac{h_T}{b_V}=0.16$, the intracardial vortex extends over the left five trabeculae, with an intertrabecular vortex only in last valley between trabeculae on the right side. For $Re\geq 50$, there is the formation of a large intracardial vortex extending throughout the cavity. However, both the trabeculae heights and $Re$ are large enough that this vortex does not wrap around each trabeculae, and intertrabecular vortices are able to form.

\begin{figure}
\centering
\includegraphics[width=0.925\textwidth]{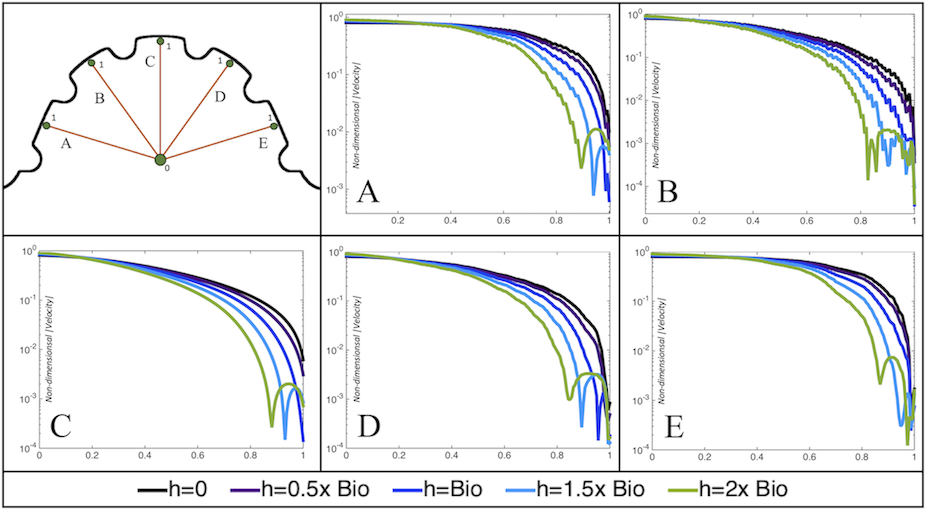}
\caption{Non-dimensional magnitude of velocity for the $Re=1$ case quantified along a line from the intracardial center (labeled ``0") and extending to the ventricular lining (labeled ``1") for various intertrabecular regions and trabeculae heights. The velocity magnitude strictly decreases from the center until around the neighboring trabeculae heights. The velocity magnitude then increases towards the center of the intratrabecular region, in some cases an order of magnitude, before dropping towards zero at the ventricle lining.}
\label{fig:Ideal_Steady_Re1_Velocity}
\end{figure}

Furthermore for the biologically relevant case of $Re=1$, Figure \ref{fig:Ideal_Steady_Re1_Velocity}, illustrates the magnitude of velocity from the intradcardial center to the ventricular lining for various intertrabecular regions and trabeculae heights. It is clear that for larger trabeculae heights, the velocity decays moving away from the intracardial center at a faster rate  than those cases with shorter or no trabeculae. However, in some cases when the trabeculae height is approximately the biologically relevant height or larger,  a local minima in velocity magnitude occurs at a distance from the bottom of the intratrabecular region that is approximately equal to the height of the trabeculae. As one continues towards the cardiac wall, the magnitude of velocity increases in the middle of the intratrabecular region and then approaches zero. This supports that there is intertrabecular vortex formation in these cases, as shown in Figure \ref{SteadyFlow_Results}. The velocities measured in such intertrabecular regions are less than those shown in the cases with no or smaller trabeculae. 

\begin{figure}
\centering
\includegraphics[width=0.925\textwidth]{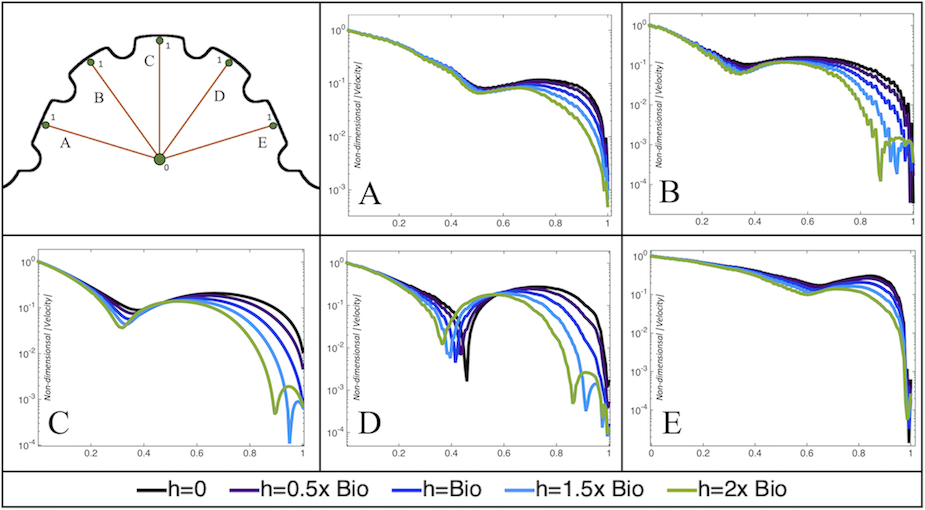}
\caption{Non-dimensional magnitude of velocity measurements for the $Re=100$ case quantified along a line from the intracardial center (labeled ``0") and extending to the ventricular lining (labeled ``1") for various intertrabecular regions and trabeculae heights. The velocity magnitude decreases from the center and then then increases before decreasing again as one approaches a distance from the wall that is equal to the neighboring trabeculae heights. As one moves between the trabeculae, the velocity magnitude again increases, in some cases an order of magnitude, before dropping towards zero at the ventricle lining.}
\label{fig:Ideal_Steady_Re100_Velocity}
\end{figure}

Interestingly, similar results are seen in the $Re=10$ case, see Figure \ref{fig:Ideal_Steady_Re10_Velocity} in Appendix \ref{app:ideal_steady_inflow}. However in the case of $Re=100$, slightly different quantitative behavior is observed, see Figure \ref{fig:Ideal_Steady_Re100_Velocity}. Note that the local minima in velocity magnitude is still observed at the neighboring trabeculae height away from the ventricular wall, followed by an increase and then decrease as one moves closer to the wall. In contrast, as you measure away from the intracardial center, the velocity magnitude decreases (as before), but it also increases and then decreases before reaching the trabeculae height distance from the ventricle wall. This is due to the presence of a large intracardial vortex that forms, as illustrated in Figure \ref{SteadyFlow_Results}.

%
%
%
%
%
%
%
%

\subsection{Pulsatile Flow through Idealized Trabeculated Chambers}


Next we consider the same idealized trabeculated ventricle but use a pulsatile inflow condition, as described in Appendix \ref{app:IB_Method}. The pulsation frequency is given by a dimensionless frequency close to that reported for a 4 \textit{dpf} embryonic zebrafish ($\tilde{f}=1.0$). The $Re$ was set to 0.1, 1.0, 10 and 100. The dimensionless trabecular heights, $\frac{h_T}{b_V}$ were varied from 0.0 to 0.16. Recall that the biologically relevant $Re$ is about one, and the biologically relevant dimensionless trabecular height is about 0.08. Snapshots of the streamlines showing the flow patterns are taken from the last pulse for each simulation using either $9$ or $10$ time points. 

\begin{figure}
\centering
\includegraphics[width=0.925\textwidth]{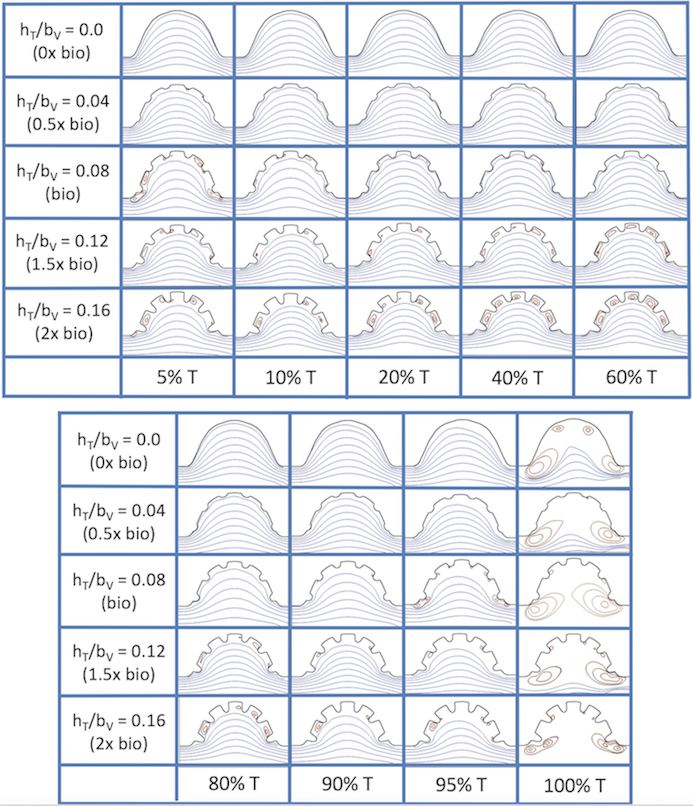}
\caption{Streamline analysis performed for the case of pulsatile flow into the trabeculated ventricle of a zebrafish at 4 \textit{dpf} for $Re=0.1$ and varying trabeculae heights.}
\label{pulse_Re0pt1_Results}
\end{figure}

\begin{figure}
\centering
\includegraphics[width=0.925\textwidth]{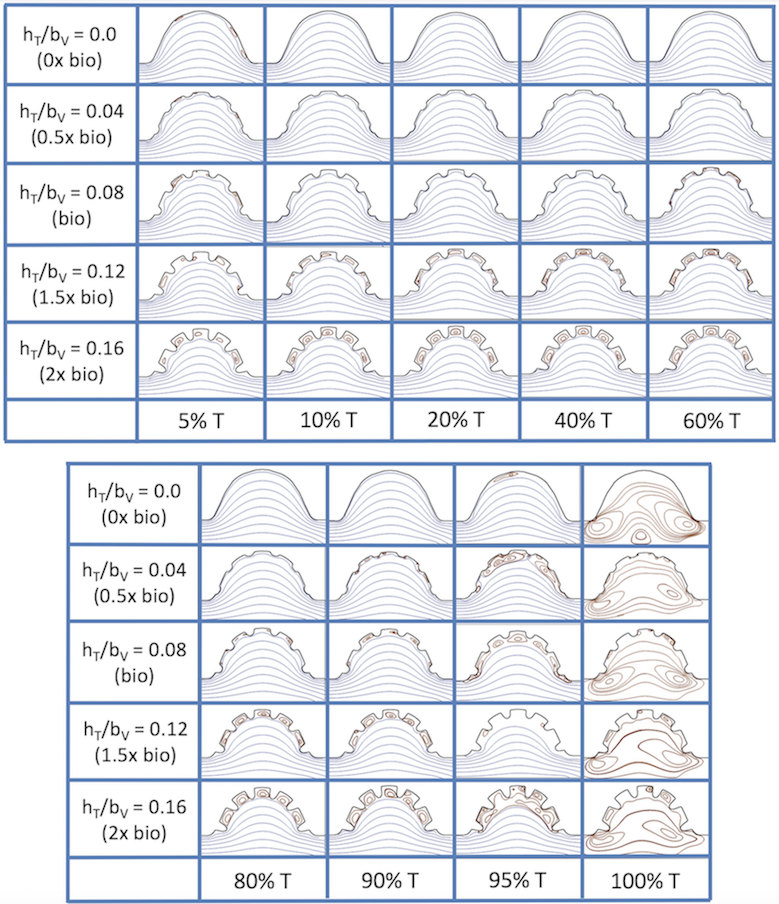}
\caption{Streamline analysis performed for the case of pulsatile flow into the trabeculated ventricle of a zebrafish at 4 \textit{dpf} for $Re=1.0$ and varying trabeculae heights.}
\label{pulse_Re1_Results}
\end{figure}

Figures \ref{pulse_Re0pt1_Results} and \ref{pulse_Re1_Results} show streamline plots taken at 9 snapshots in time for lower $Re$ cases, $Re = 0.1, 1.0$, respectively. The streamlines are shown for 5\%, 10\%, 20\%, 40\%, 50\%, 80\%, 90\%, 95\%, and 100\% of the pulse. Finer increments in time are given towards the beginning and end of the pulse to illustrate the rapidly changing dynamics. The $Re=0.1$ and $1.0$ cases show similar results. For the majority of the pulse, the flow moves smoothly from left to right within the ventricle. In between the trabeculae, vortices form during most of the pulse if the dimensionless trabecular height is at least 0.08. The development of these vortices causes the flow near the endothelial cells to move from right to left between the trabeculae and from left to right on the top of the trabeculae. In most cases, transient vortices form as the flow is decelerated at the end of the pulse. For $Re = 1.0$, intertrabecular vortices form for small trabeculae, $\frac{h_T}{b_V}=0.04$, as the flow decelerates. This is different than in the steady inflow counterpart case in Section \ref{sec:ideal_steady}, where vortex formation only happened in for biologically relevant heights or greater.  

\begin{figure}
\centering
\includegraphics[width=0.925\textwidth]{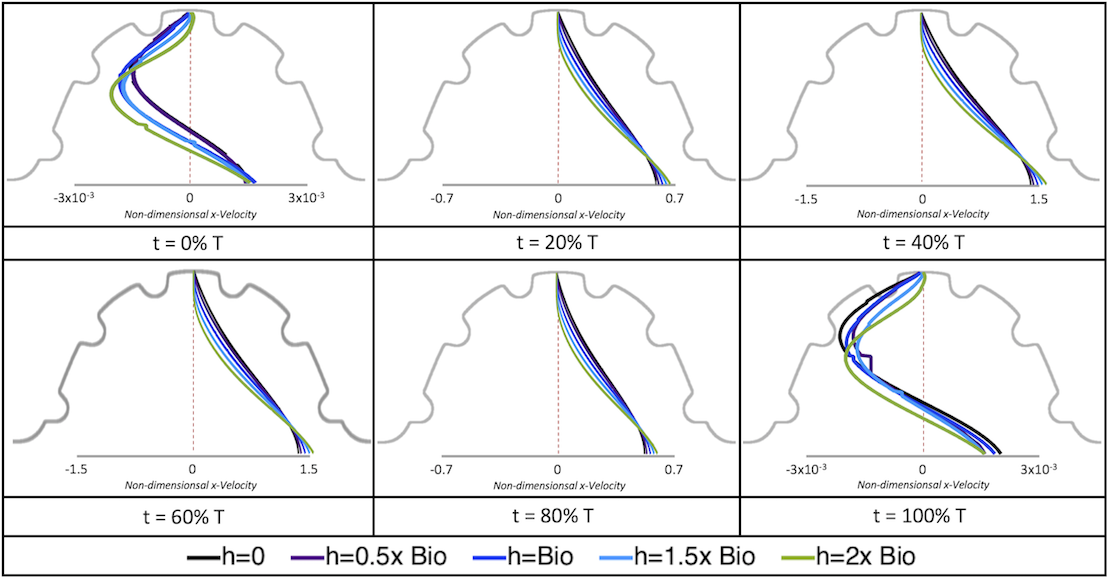}
\caption{Snapshots during a pulsation cycle in the $Re=1$ case of the horizontal velocity measured from the intracardial center to the intertrabecular region directly above for multiple trabeculae heights.}
\label{fig:pulse_Re1_Velocity_Results}
\end{figure}

Next we considered the horizontal velocity extending from the intracardial center to the intertrabecular region directly above it in the case of $Re=1$. These results are shown in Figure \ref{fig:pulse_Re1_Velocity_Results}. Flow velocities are least near the ventricular lining, with magnitude of velocities decreasing at a slightly accelerated rate for larger trabeculae. On the other hand, at the intracardial center, the horizontal velocity increases with larger trabeculae in the middle of a pulsation cycle. As the pulsation cycle resides, it is clear that the flow direction also changes, giving rise to an intracardial vortex, as detailed in Figure \ref{pulse_Re1_Results}. 

\begin{figure}
\centering
\includegraphics[width=0.925\textwidth]{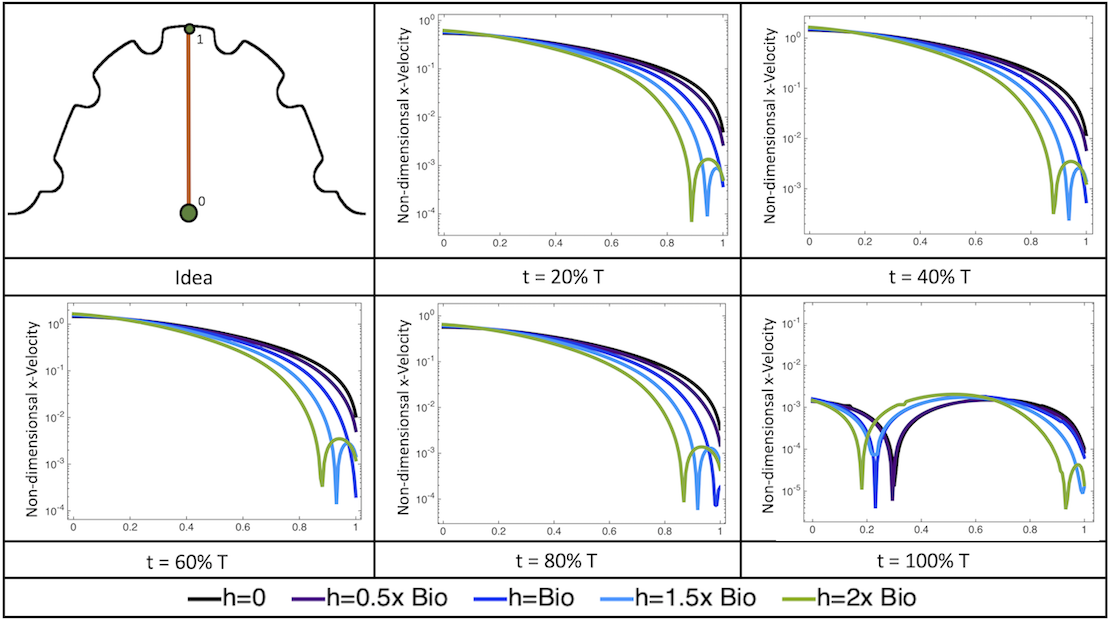}
\caption{Snapshots during a pulsation cycle in the $Re=1$ case of the horizontal velocity measured from the intracardial center (labeled ``0") to the intertrabecular region (labeled ``1") directly above for multiple trabeculae heights. The trabeculae cause a drop in velocity as one nears the ventricular wall.}
\label{fig:pulse_Re1_Velocity_Data}
\end{figure}

Figure \ref{fig:pulse_Re1_Velocity_Data} shows the magnitude of velocity as as one moves along a line drawn from the top of the ventricle to the base of the middle intratrabecular region for five trabeculae heights at five times during the pulse cycle. Note that during a pulsation cycle, the snapshots taken during the middle of the pulse are similar to that of the steady inflow case for $Re=1$ shown in Figure \ref{SteadyFlow_Results}. For larger trabeculae heights, the horizontal velocity reaches a minimum at distance that is about a trabecular height away from the ventricle wall. As one moves between the trabeculae, the magnitude of the velocity increases and then approaches as one nears the wall. These velocity profiles confirm the formation of vortices as shown in Figure \ref{pulse_Re1_Results}. 

\begin{figure}
\centering
\includegraphics[width=0.925\textwidth]{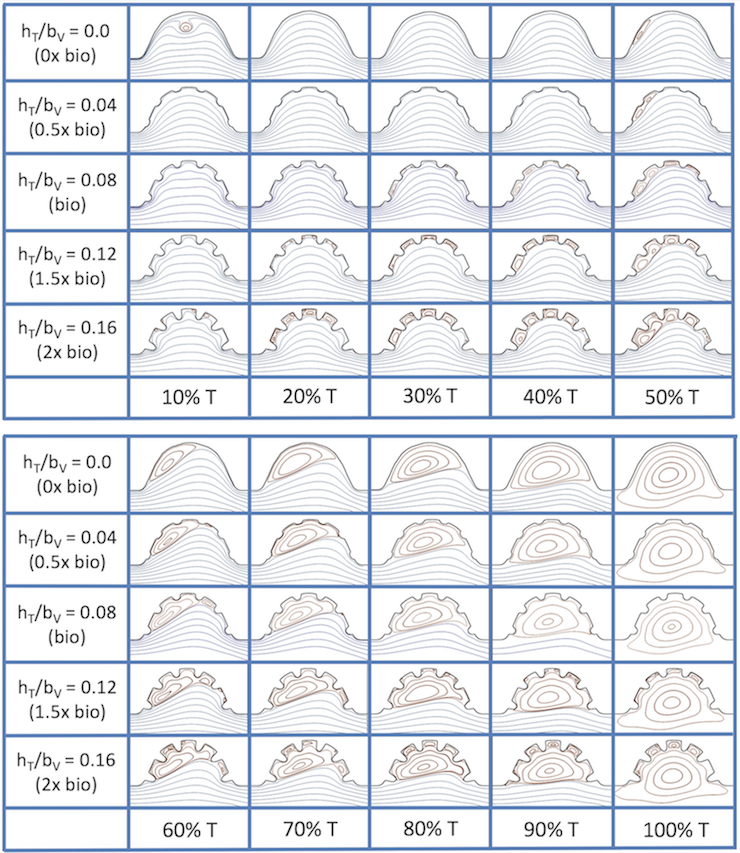}
\caption{Streamline analysis performed for the case of pulsatile flow into the idealized trabeculated ventricle of a zebrafish at 4 \textit{dpf} for $Re=10.0$ and varying trabeculae heights.}
\label{pulse_Re10_Results}
\end{figure}

Figure \ref{pulse_Re10_Results} shows streamline plots for $Re = 10$ at ten evenly spaced times during a pulse. Intertrabecular vortices form during the first half of the pulse if the dimensionless trabecular height is at least 0.04. For all geometries, intracardial vortices form during the last half of the pulse. The formation of the intracardial vortex annihilates the intertrabecular vortices, at least initially. The intracardial vortices form on the upstream side of the chamber, and grow to fill the entire chamber by the end of the pulse. The intertrabecular vortices form again towards the end of the pulse for $\frac{h_T}{b_V} =0.12, 0.16$. Note that the presence of the intracardial vortex causes the intertrabecular vortices to change direction so that they spin clockwise (and the intracardial vortices spin counterclockwise).

\begin{figure}
\centering
\includegraphics[width=0.925\textwidth]{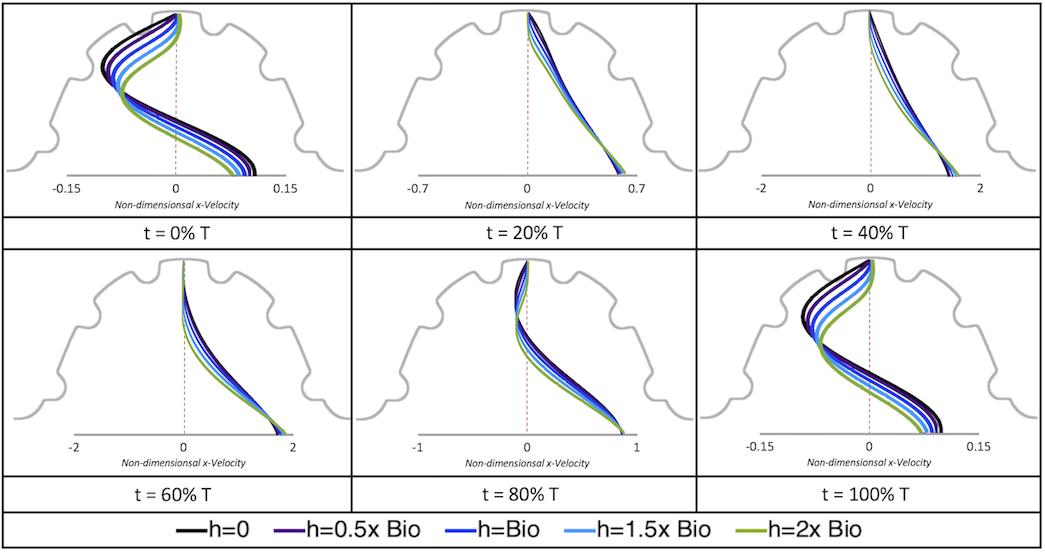}
\caption{Snapshots during a pulsation cycle in the $Re=10$ case of the horizontal velocity measured from the intracardial center to the intertrabecular region directly above for multiple trabeculae heights.}
\label{fig:pulse_Re10_Velocity_Results}
\end{figure}

We also considered the horizontal velocity extending from the intracardial center to the intertrabecular region directly above for $Re=10$, as shown in Figure  \ref{fig:pulse_Re10_Velocity_Results}. Similarly quantitative behavior is seen as in the $Re=1$ case, see Figure \ref{fig:pulse_Re1_Velocity_Results}; there is still significantly less flow in the intertrabecular region. However, the flow velocities are significantly different in the cavity below the intertrabecular region because of the presence of an intracardial vortex. It is clear there is flow reversal given the differences in sign of the horizontal velocity.

\begin{figure}
\centering
\includegraphics[width=0.925\textwidth]{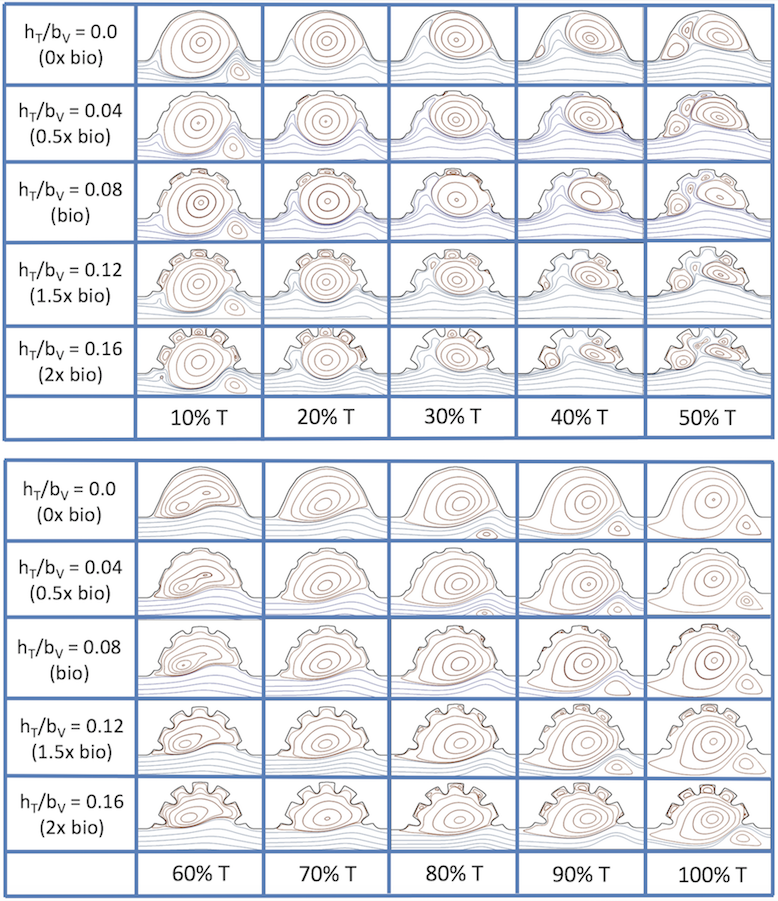}
\caption{Streamline analysis performed for the case of pulsatile flow into the trabeculated ventricle of a zebrafish at 4 \textit{dpf} for $Re=100$ and varying trabeculae heights.}
\label{pulse_Re100_Results}
\end{figure}

The results of the inertial dominated case, $Re=100$, are shown in Figure \ref{pulse_Re100_Results}. In all cases, a large intracardial vortex that fills the entire chamber is observed at the end of the pulse and beginning of the next pulse. As the flow accelerates, the intracardial vortex is pushed downstream, and another intracardial vortex begins to form ($t = 0.4 T - 0.5 T$). One or more oppositely spinning vortices form between the trabeculae or between the two counterclockwise spinning intracardial vortices when $t = 0.5 T$. The upstream intracardial vortex combines with the original intracardial vortex such that one large intracardial vortex is observed around $t = 0.7 T$. When this occurs, the oppositely spinning vortices are annihilated. For $\frac{h_T}{b_V}\geq 0.08$, oppositely spinning intertrabecular vortices reappear at the end of the pulse.


\begin{figure}
\centering
\includegraphics[width=0.925\textwidth]{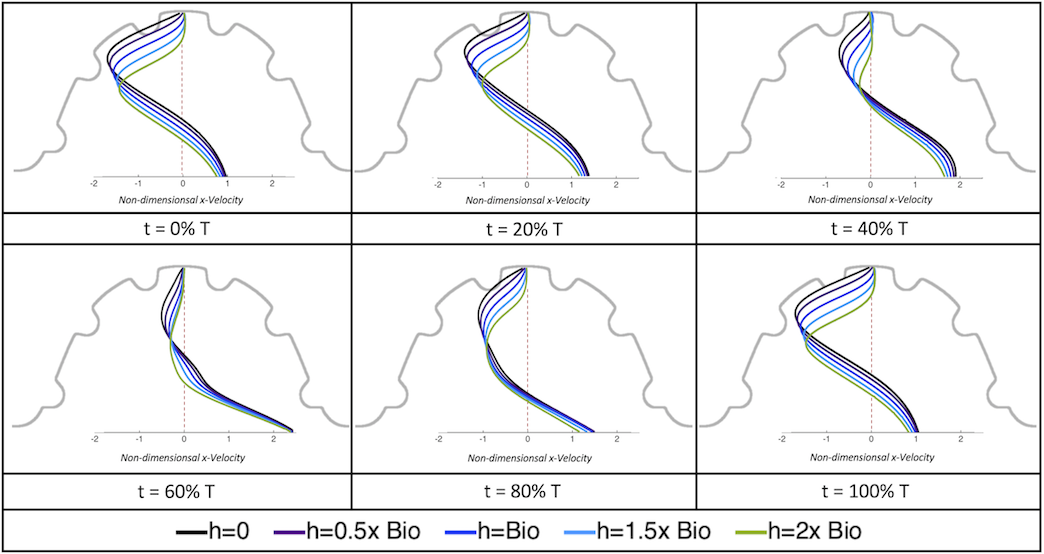}
\caption{Snapshots during a pulsation cycle in the $Re=100$ case of the horizontal velocity measured from the intracardial center to the intertrabecular region directly above for multiple trabeculae heights.}
\label{fig:pulse_Re100_Velocity_Results}
\end{figure}

Due to the formation of large intracardial vortices, the horizontal velocity changes sign when measured from the center of the cavity and proceed directly upward toward the ventricle lining, see Figure \ref{fig:pulse_Re100_Velocity_Results}. Similar quantitative behavior is seen as in the $Re=10$ case. A substantial difference is the presence of a intracardial vortex that remains largely throughout the pulsation cycle. Moreover, similar to the other cases of $Re=1,10$, the velocity is significantly decreased within the intertrabecular region, even for $Re=100$. 

%
%

%
%

\section{Conclusions}

Two-dimensional immersed boundary simulations were used to solve for the flow fields within both biologically realistic geometries (3 \textit{dpf}, 80 \textit{hpf}, and 5 \textit{dpf} WT zebrafish and a 7 \textit{dpf ErbB2}-inhibited zebrafish from \cite{Liu:2010}) and idealized models of trabeculated ventricles. Specifically, we investigated the intracardial and intertrabecular fluid dynamics searching for possible vortex formation and spatially-varying velocity gradients. 

This work focused specifically on the presence or absence of vortices given their significance to both the magnitude and direction of flow as well as the mixing patterns within the ventricle. When an intracardial vortex forms, the direction of the flow changes. When an intracardial vortex forms in unsteady flow, the direction of flow can change during the beat cycle, and the stagnation point moves along the cardiac wall. Since endothelial cells are known to sense and respond to changes in both magnitude and direction of flow, the formation and motion of these vortices could be important epigenetic signals. 

The simulations revealed unexpected complexities in vortex dynamics as bulk flow moves through the chamber. In the cases of biologically realistic geometries, no large intracardial vortices developed at the biologically realistic fluid scale ($Re\sim1$) in the steady inflow case; in the pulsatile case a large intracardial vortex formed when inflow velocities were minimal between pulses. However, for both steady and pulsatile inflow intertrabecular vortices formed in pronounced trabeculated regions (Figures \ref{fig:real_Steady} and \ref{fig:real_Pulsatile}). While similar behavior was observed in both the steady and pulsatile inflow cases, more vortices formed in the unsteady flow cases in regions with less pronounced trabeculation. These results were consistent with the idealized trabeculation model as well (Figures \ref{SteadyFlow_Results} and \ref{pulse_Re1_Results}). When vortices form in intertrabecular regions, the flow changes direction in those areas compared to the direction of bulk flow in the chamber. In instances cases where not all intertrabecular spaces have a vortex, the flow between different trabeculae will move in different directions.

Also as expected the velocity tapers off when measuring flow speeds close to the ventricular lining. In most cases the flow tapers off approximately three-orders of magnitude before reaching an intertrabecular region. Interestingly, in these regions where there is pronounced vortex formation, the velocity increases and then decreases as it nears the wall (for the pulsatile cases with realistic geometries see Figures \ref{fig:real_Pulsatile_3dpf_LoguMag_Data}, \ref{fig:real_Pulsatile_5dpf_LoguMag_Data}, \ref{fig:real_Pulsatile_7dpf_ErbB2_LoguMag_Data}. Note that the corresponding steady inflow cases show similar trends). In the 5 \textit{dpf} WT case the velocity may increase an order of magnitude during portions of a pulsation cycle (see Figure \ref{fig:real_Pulsatile_3dpf_LoguMag_Data}) while in other cases the increase is modest, such as in the 3 \textit{dpf} WT case (see Figure \ref{fig:real_Pulsatile_5dpf_LoguMag_Data}). These results are quantitatively similar to those of the idealized model case at $Re=1$ for both pulsatile and steady inflow. The presence of trabeculae appear to control the fluid velocities in this regions, which helps govern the amount of shear-stress felt at the endothelial layer.


The idealized model cases expanded the study by allowing us to easily manipulate the system to understand its sensitivity to vortex formation, due to its complex geometry, fluid scale, and inflow characteristics. In this vein, we increased the span of our larger fluid scale from $Re=0.1$ to $Re=100$, rather than simply the biological case of $Re=1$ for a 4 \textit{dpf} embryonic zebrafish heart. We also investigated equally sized idealized trabeculae in a chamber, and varied the heights from no trabeculae to trabeculae twice as large as biologically relevant according to Figure \ref{Model_Geometry}a, and included both pulsatile and steady inflow simulations to parse effects of unsteady flows on vortex formation. 

Moreover, diversity of trabeculated hearts across the animal kingdom are far and wide, and may cover a large spectrum of length and fluid ($Re$) scales. Even some invertebrate hearts contain ventricular trabeculation. In those invertebrates, their heart's morphology resembles that of lower vertebrates with sedentary lifestyles \cite{Sedmera:2010,Burggren:1988,Burggren:1994}. Some anthropods and mollusks contain ventiruclar trabeculae, such as blue crabs (\textit{Callinectes sapidus}) \cite{McGaw:2002}, bar clams (\textit{Spisula solidissima}) \cite{Collis:2006}, oysters \cite{Hu:2000}, snails \cite{Kodirov:2011}, as well as  octopus and squids \cite{Wilbur:2013}.  However, any quantitative measurements detailing trabeculae morphology and flow measurements at varying time-points are unknown. It is possible that in these hearts the idealized simulations at higher $Re$ or for larger trabecular heights are relevant to these invertebrate hearts.

The idealized models showed that a large intracardial vortex forms around $Re \approx 20$ when steady flow is pushed through the chamber, while a similar sized vortex forms for $Re=10$ when the flow is pulsatile. In general, pulsatile flow lowers the $Re$ and trabeculae height needed to generate vortices. For both steady and unsteady flows as the trabeculae grow into the chamber, another bifurcation occurs in which small vortices form between each trabecula. Depending upon the $Re$ and the morphology, the intertrabecular vortices can form without the presence of a large intracardial vortex, see Figure \ref{SteadyFlow_Results} for steady cases or Figures \ref{pulse_Re0pt1_Results} or \ref{pulse_Re1_Results} for unsteady flow cases of $Re=0.1$ or $1.0$, respectively. In other cases, typically at higher $Re$, both the intracardial and intertrabecular vortices form, see Figures \ref{pulse_Re10_Results} or \ref{pulse_Re100_Results}, for unsteady inflow for $Re=10$ or $100$, respectively. In all corresponding cases, the presence of large intracardial vortices changes the direction of the intertrabecular vortices. Note that in the biologically relevant case of $Re=1$ intracardial vortices do not form; this is consistent with the biologically accurate geometries as well.

Importantly, the idealized study demonstrates that small changes in viscosity, scale, morphology, and contraction dynamics can substantially influence bulk flow properties in the embryonic heart. This presents an interesting challenge since each of these parameters is continuously changing during development. In addition, estimating the effective viscosity of the embryonic blood is nontrivial. Given the sensitivity of the flow to such small perturbations, it is necessary to use well resolved numerical grids that are experimentally validated.

It is evident that there is a strongly coupled relationship between intracardial hemodynamics, genetic regulatory networks, and cardiac conduction. Besides contractions of the myocardial cells, which in turn drive blood flow, hemodynamics are directly involved in proper pacemaker and cardiac conduction tissue formation \cite{Tucker:1988}. Also, shear stresses are found to govern the conduction velocity distribution of action potentials within the myocardium \cite{Reckova:2003}. Any changes in the emrbyonic heart's conduction properties will also affect the intracardial shear stresses, pressures, patterns of cyclic strains, and advection of morphogens. It is indeed a chicken and the egg scenario, especially when considering the first experiments that saw the importance of fluid dynamics in heart morphogenesis were performed in chicken embryos \cite{Chapman:1918}. Dedicated initiatives to decipher exact cellular signalling pathways and genetic regulatory networks may be able to help further parse the causes of cardiac dysfunction. While CFD provides a robust framework to extract cardiac flow information (fluid velocities, shear stress distributions, pressures, etc.), coupling this data into a multi-scale cellular model is imperative to better understand the causes of many congenital heart diseases. 

Unfortunately the exact mechanisms of mechanotransduction are not yet clearly understood \cite{Weinbaum:2003,Paluch:2015}. Biochemical signals are thought to propagate throughout a pipeline of epigenetic signaling mechanisms, which may regulate of gene expression, cellular differentiation, proliferation, and migration \cite{Chen:2014}. \textit{In vitro} studies have discovered that endothelial cells detect shear stresses as low as $0.2\ dyn/cm^2$, resulting in up or down regulation of gene expressions \cite{Olesen:1988}. Shear stresses around $\sim8-15\ dyn/cm^2$ are known to cause cytoskeletal rearrangement \cite{Davies:1986}. These aforementioned shear stresses are well in the range of those measured within emrbyonic hearts, $\sim2\ dyn/cm^2$ and $\sim75\ dyn/cm^2$ at approximately $1.5$ and $4.5$ \textit{dpf}, respectively \cite{Hove:2003}. Mapping out the connection between fluid dynamics, the resulting ventricular stresses, electrophysiology, and the mechanical regulation of developmental regulatory networks are paramount to moving towards a more holistic understanding of heart development.

\vspace{6pt} 



%
%

\authorcontributions{Conceptualization, N.A.B and L.A.M.; Methodology and Software, N.A.B., D.R.D., and L.A.M.; Validation, N.A.B., D.R.D., A.N.L., and L.A.M..; Formal Analysis, Investigation, and Data Curation, N.A.B. and L.A.M; Writing, Original Draft Preparation, N.A.B., A.N.L., and L.A.M.; Writing, Review \& Editing, N.A.B., A.N.L., L.A.S., J.L., and L.A.M.; Visualization, N.A.B.; Funding Acquisition, N.A.B. and L.A.M.}

%
%

\funding{This project was funded by NSF DMS CAREER \#1151478 awarded to L.A.M. Funding for N.A.B. and L.A.S. was provided from an National Institutes of Health T32 grant [HL069768-14; PI, Nobuyo Maeda and Christopher Mack] and the Support of Scholarly Activities Grant (TCNJ).}

%
%

\acknowledgments{The authors would like to thank Steven Vogel for conversations on scaling in various hearts. We would also like to thank Lindsay Waldrop, Austin Baird, and William Kier for discussions on embryonic hearts.}

\conflictsofinterest{The authors declare no conflict of interest.} 

\abbreviations{The following abbreviations are used in this manuscript:\\

\noindent 
\begin{tabular}{@{}ll}
DPF & Days Post Fertilization \\
HPF & Hours Post Fertilization \\
WT & Wild type embryo \\
CFD & Computational Fluid Dynamics \\
Re & Reynolds Number \\
IB & Immersed Boundary Method 
\end{tabular}}

%
%

\appendixtitles{yes} 
\appendixsections{multiple} 

\appendix
%
%

\section{Immersed Boundary Method}
\label{app:IB_Method}


The immersed boundary method  \cite{Peskin:2002} was used to solve for the flow velocities within the geometric model from Section \ref{Geometry}. The immersed boundary method (IB) has been successfully used to study the fluid dynamics of a variety of biological problems in the intermediate Reynolds number range, defined here as $0.01<Re<1000$ (see, for example, \cite{Jung:2001,Hershlag:2011,Bhalla:2013a,Tytell:2010}). Although, IB is capable of solving fully coupled fluid-structure interaction systems, here we only use it to solve fluid flow through complex model geometry. The model consists of stiff boundaries that are immersed within an incompressible fluid of dynamic viscosity, $\mu$, and density, $\rho$. The fluid motion is described using the full 2D Navier-Stokes equations in Eulerian form, given a

\begin{equation}
\label{Navier_Stokes} \rho \left( \frac{\partial {\bf{u}}({\bf{x}},t) }{\partial t} + {\bf{u}}({\bf{x}},t)\cdot \nabla {\bf{u}}({\bf{x}},t) \right) = -\nabla p({\bf{x}},t) + \mu \Delta {\bf{u}}({\bf{x}},t) + {\bf{F}}({\bf{x}},t) \\
\end{equation}
\begin{equation}
\label{Incompressibility} \nabla\cdot {\bf{u}}({\bf{x}},t) = 0,
\end{equation}
where ${\bf{u}}({\bf{x}},t) = (u({\bf{x}},t),v({\bf{x}},t))$ is the fluid velocity, $p(\bf{x},t)$ is the pressure, ${\bf{F}}({\bf{x}},t)$ is the force per unit volume (area in $2D$) applied to the fluid by the immersed boundary, i.e., the ventricle geometry. The independent variables are the position, ${{\bf{x}}}= (x,y)$, and time, $t$. Eq.(\ref{Navier_Stokes}) is equivalent to the conservation of momentum for a fluid, while Eq.(\ref{Incompressibility}) is a condition mandating that the fluid is incompressible. \\

The interaction equations between the fluid and the immersed structure are given by
\begin{equation}
\label{IBM_Force} {\bf{F}}({\bf{x}},t) = \int {\bf{f}}(r,t)\delta({\bf{x}}-{\bf{X}}(r,t)) dr
\end{equation}
\begin{equation}
\label{IBM_Velocity} {\bf{U}}({\bf{X}}(r,t),t) = \frac{\partial {\bf{X}}(r,t)}{\partial t} = \int {\bf{u}}({\bf{x}},t) \delta( {\bf{x}} - {\bf{X}}(r,t) ) d{\bf{x}},
\end{equation}
where ${\bf{X}}(r,t)$ gives the Cartesian coordinates at time $t$ of the material point labeled by Lagrangian parameter $r$, ${\bf{f}}(r,t)$ is the force per unit area imposed onto the fluid by elastic deformations in the boundary, as a function of the Lagrangian position, $r$, and time, $t$. Eq.(\ref{IBM_Force}) applies a force from the immersed boundary to the fluid grid through a delta-kernel integral transformation. Eq.(\ref{IBM_Velocity}) sets the velocity of the boundary equal to the local fluid velocity.

The force equations are specific to the application. In a simple case where a preferred motion or position is enforced, boundary points are tethered to target points via springs. The equation describing the force applied to the fluid by the boundary in Lagrangian coordinates is given by ${\bf{f}}(r,t)$ and is explicitly written as,
\begin{equation}
\label{IBM_force_explicit} {\bf{f}}(r,t) = k_{target} \left( {\bf{Y}}(r,t) - {\bf{X}}(r,t) \right),
\end{equation}
where $k_{target}$ is the stiffness coefficient, and $\bf{Y}(r,t)$ is the prescribed Lagrangian position of the target structure. In all simulations the immersed structure was held nearly rigid by applying a force proportional to the distance between the location of the actual boundary and the preferred position. The deviation between the actual and preferred positions can be controlled with the variable $k_{target}$.\\


The fluid flow is driven through the immersed boundary using either pulsatile parabolic inflows or a linear ramp to steady parabolic inflow at the location of the AV canal. The equations describing the specific inflow boundary conditions are given in Table(\ref{BCs}). A partial Neumann outflow condition is enforced in the direction of flow at the outlet. This outflow condition is given as
\begin{equation}
 \label{BC_outflow}\left( \begin{array}{c}
 \frac{\partial u}{\partial \hat{n}} \\
v \end{array}\right) =  \left( \begin{array}{c}
0 \\
0 \end{array} \right).
\end{equation}
\noindent where $u$ and $v$ are the $x-$ and $y-$components of the fluid velocity, respectively, and $\frac{\partial u}{\partial n}$ is the directional derivative of the $x-$component of the velocity taken in the direction normal to the boundary of the fluid domain.
\begin{table}
\centering
\begin{tabular}{c|c}
\hline
Case & Inflow BC \\
\hline
Steady Inflow & $\bf{u}_{in} = \left( \begin{array}{c}
\frac{V_{in}}{d_{AV}^2} \tanh(2t) \left( \frac{1}{4 d_{AV}^2} - y^2 \right) \\
0 \end{array} \right)$ \\ 
Pulsatile Inflow & $\bf{u}_{in} = \left( \begin{array}{c}
\frac{V_{in}}{d_{AV}^2} |\sin(2\pi f t)| \left( \frac{1}{4 d_{AV}^2} - y^2 \right) \\
0 \end{array} \right)$ \\
\hline
\end{tabular}
\caption{Inflow boundary conditions for both simulations, one pertaining to parabolic steady inflow and the other corresponding to a parabolic pulsatile inflow. The parameters used for the boundary conditions are $f$, the non-dimensional frequency, which is matched to the zebrafish heart at $96\ hpf$, and $V_{in}$, the maximum inflow velocity.}
\label{BCs}
\end{table}


We used an adaptive and parallelized version of the immersed boundary method, IBAMR \citep{BGriffithIBAMR,Griffith:2007}, to perform the simulations involving the highly idealized trabecular models, for both the steady inflow and pulsatile inflow cases. IBAMR is a C++ framework that provides discretization and solver infrastructure for partial differential equations on block-structured locally refined Eulerian grids \citep{MJBerger84,MJBerger89} and on Lagrangian (structural) meshes. IBAMR also includes infrastructure for coupling Eulerian and Lagrangian representations.  

The Eulerian grid on which the Navier–Stokes equations were solved was locally refined near the immersed boundaries and regions of vorticity with a threshold of $|\omega| > 0.05$. This Cartesian grid was organized as a hierarchy of four nested grid levels, and the finest grid was assigned a spatial step size of $dx = D/1024$, where $D$ is the length of the domain. The ratio of the spatial step size on each grid relative to the next coarsest grid was 1:4. The temporal resolution was varied to ensure stability. Each Lagrangian point of the immersed structure was chosen to be $\frac{D}{2048}$ apart (twice the resolution of the finest fluid grid).


%
%

\section{Methods: Biologically Realistic Geometries}
\label{app:bio_geo}

We present two more computational geometries we will analyze for WT zebrafish embryos at 80 \textit{hpf} and 5 \textit{dpf}. Note that the 80 \textit{hpf} is close to the 3 \textit{dpf} time-point and the 5 \textit{dpf} zebrafish illustrates natural variation in WT zebrafish from the embryo given in Figure \ref{Bio_Geometry}.

\begin{figure}[h!]
\centering
\includegraphics[width=0.925\textwidth]{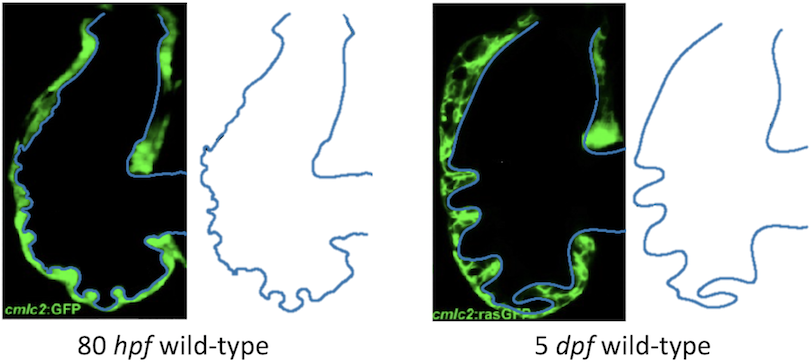}
\caption{Extracting biologically realistic geometries of trabeculated geometries at different time-points from \cite{Liu:2010}. Geometries are given for a WT zebrafish at 80 \textit{hpf} and 5 \textit{dpf}.}
\label{Bio_Geometry_App}
\end{figure}

%
%

\section{Results: pulsatile inflow for biologically realistic geometries results}
\label{app:bio_results}

\begin{figure}[h!]
\centering
\includegraphics[width=0.975\textwidth]{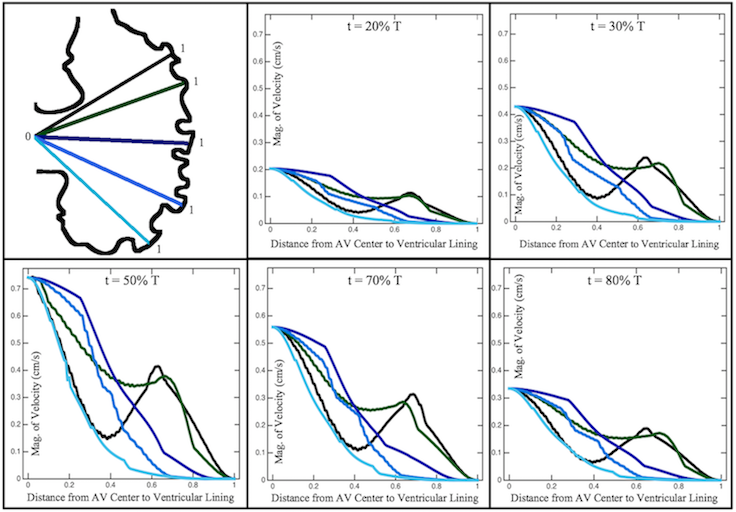}
\caption{Plots showing how the magnitude of velocity decays from the center of the AV canal to the ventricle wall for four different lines across the chamber using a 5 \textit{dpf} WT embryo's geometry.}
\label{fig:real_Pulsatile_3dpf_uMag_Data}
\end{figure}

\begin{figure}[h!]
\centering
\includegraphics[width=0.975\textwidth]{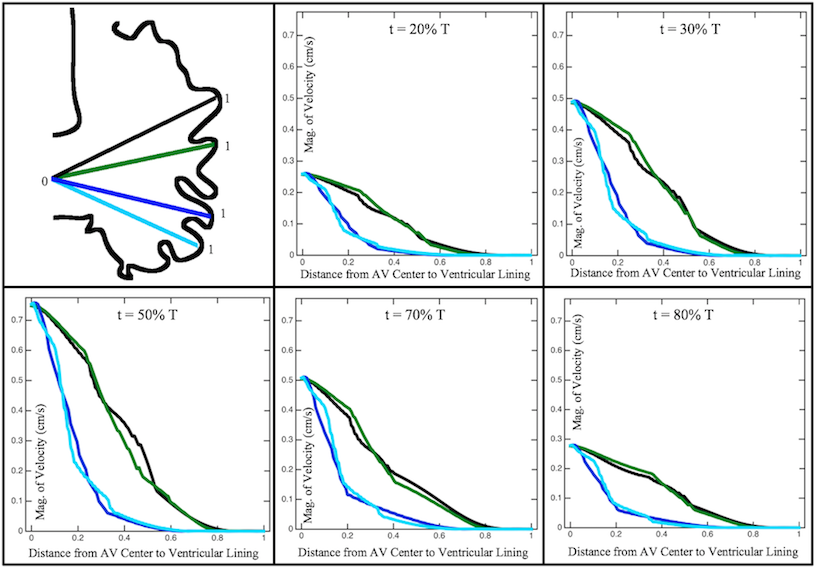}
\caption{Plots showing how the magnitude of velocity decays from the center of the AV canal to the ventricle wall for four different lines across the chamber using a 5 \textit{dpf} WT embryo's geometry.}
\label{fig:real_Pulsatile_5dpf_uMag_Data}
\end{figure}

\begin{figure}[h!]
\centering
\includegraphics[width=0.975\textwidth]{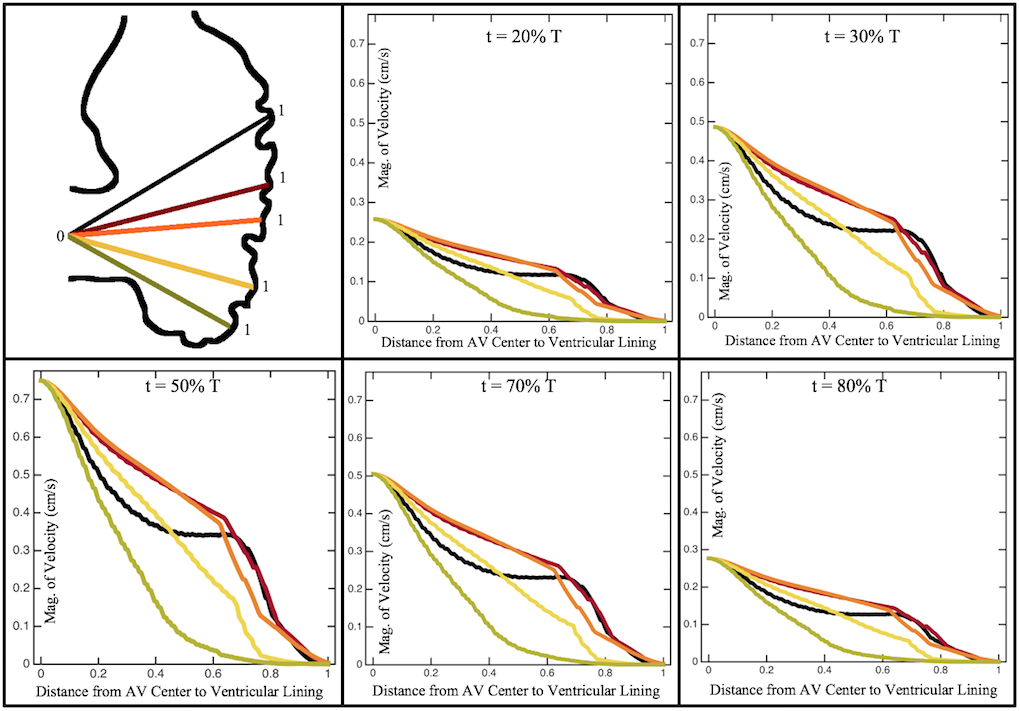}
\caption{Plots showing how the magnitude of velocity decays from the center of the AV canal to the ventricle wall for five different lines across the chamber using the \textit{ErbB2}-inhibited emrbyo's geometry.}
\label{fig:real_Pulsatile_ErbB2_7dpf_uMag_Data}
\end{figure}

Here we give more velocity measurements extending from the AV canal center to intertrabecular regions along the ventricular canal for cases of a 3 and 5 \textit{dpf} WT and 7 \textit{dpf ErbB2} inhibited zebrafish, Figures \ref{fig:real_Pulsatile_3dpf_uMag_Data}, \ref{fig:real_Pulsatile_5dpf_uMag_Data}, and \ref{fig:real_Pulsatile_ErbB2_7dpf_uMag_Data}, respectively. These figures are given to compare to their cases of steady inflow. Note that in all cases, the velocity profiles taken at $50\%$ of the pulsation cycle looks qualitatively identical to its corresponding steady inflow case. Note that this data was presented in Figures \ref{fig:real_Pulsatile_3dpf_LoguMag_Data}, \ref{fig:real_Pulsatile_5dpf_LoguMag_Data}, and \ref{fig:real_Pulsatile_7dpf_ErbB2_LoguMag_Data} as semi-log plots to illustrate the extent of the velocity decay as measured near the ventricular lining.

%
%

\section{Results: Steady inflow for Idealized Geometry}
\label{app:ideal_steady_inflow}

In this appendix we present magnitude of velocity results in the idealized geometry case for $Re=10$, see Figure \ref{fig:Ideal_Steady_Re10_Velocity} below. Figure \ref{fig:Ideal_Steady_Re10_Velocity} shows similar qualitative trends to that of the $Re=1$ case in Figure \ref{fig:Ideal_Steady_Re1_Velocity} of Section \ref{sec:ideal_steady}. However, differences arise in the actual quantitative magnitude of velocity measurements, where in the $Re=10$ case, the velocity magnitudes are generally slightly greater.

\begin{figure}[h!]
\centering
\includegraphics[width=0.9\textwidth]{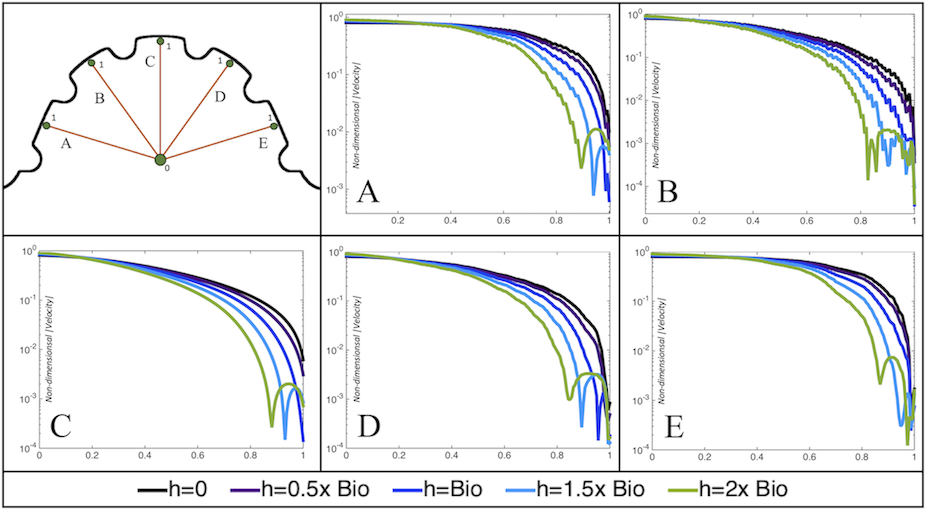}
\caption{Non-dimensional magnitude of velocity measurements for the $Re=10$ case quantified along a line from the intracardial center (labeled ``0") and extending to the ventricular lining (labeled ``1") for various intertrabecular regions and trabeculae heights. The velocity strictly decreases from the center until around the neighboring trabeculae heights, in which it the velocity increases, in some cases an order of magnitude, before dropping towards zero at the ventricle lining.}
\label{fig:Ideal_Steady_Re10_Velocity}
\end{figure}


\reftitle{References}

\bibliographystyle{mdpi}       
\bibliography{tubeheart_rbcs}   





\end{document}